\documentclass[superscriptaddress,twocolumn]{revtex4}

\usepackage{amsfonts}
\usepackage{amsmath}    
\usepackage[normalem]{ulem}
\usepackage{bm}
\usepackage{graphicx}
\usepackage{titlesec}

\newcommand{\ket}[1]{\vert#1\rangle}

\def\opone{\leavevmode\hbox{\small1\kern-3.8pt\normalsize1}}
\usepackage{color}
\titlespacing{\subsection}{1pt}{\parskip}{-\parskip}
\begin{document}
\title{Coherent storage and manipulation of broadband photons via dynamically controlled Autler-Townes splitting}

\author{Erhan Saglamyurek}
\affiliation{Department of Physics, University of Alberta Edmonton, Alberta T6G 2E1 Canada}
\author{Taras Hrushevskyi}
\affiliation{Department of Physics, University of Alberta Edmonton, Alberta T6G 2E1 Canada}
\author{Anindya Rastogi}
\affiliation{Department of Physics, University of Alberta Edmonton, Alberta T6G 2E1 Canada}
\author{Khabat Heshami}
\affiliation{National Research Council Canada, 1200 Montreal Road Ottawa, Ontario K1A 0R6, Canada}
\author{Lindsay J. LeBlanc}
\affiliation{Department of Physics, University of Alberta Edmonton, Alberta T6G 2E1 Canada}
\affiliation{Canadian Institute for Advanced Research, Toronto, Ontario M5G 1M1, Canada}

\maketitle

\textbf{The coherent control of light with matter, enabling  storage and manipulation of optical signals, was revolutionized by electromagnetically induced transparency (EIT), which is a quantum interference effect. For strong electromagnetic fields that induce a wide transparency band, this quantum interference vanishes, giving rise to the well-known phenomenon of Autler-Townes splitting (ATS). To date, it is an open question whether ATS can be directly leveraged for coherent control as more than just a case of ``bad'' EIT. Here, we establish a protocol showing that dynamically controlled absorption of light in the ATS regime mediates coherent storage and manipulation that is inherently suitable for efficient broadband quantum memory and processing devices. We experimentally demonstrate this protocol by storing and manipulating nanoseconds-long optical pulses through a collective spin state of laser-cooled Rb atoms for up to a microsecond. Furthermore, we show that our approach substantially relaxes the technical requirements intrinsic to established memory schemes, rendering it suitable for broad range of platforms with applications to quantum information processing, high-precision spectroscopy, and metrology.}

When a strong electromagnetic field resonantly drives a transition, that transition can be split into a doublet due to the dynamic, or ac, Stark effect of the field, as first reported by Autler and Townes~\cite{Autler:1955gb}.  This effect can be directly probed using a weak electromagnetic field that couples the split levels to a third level (Fig.~\ref{ATS}a-c). Since its discovery, this splitting, commonly referred to as Autler-Townes splitting (ATS), has been  observed in numerous  atomic and molecular media~\cite{Picque:1976dh,He:1992dx}, and extensively studied to describe underlying quantum optical  phenomena in laser cooling, cavity quantum electrodynamics~\cite{Zhu:1990jr,Thompson:1992jr,Bernadot:1992kg} and high resolution spectroscopy \cite{Wade:2014, Holloway:2014, Ghafoor:2014}. 

In the context of coherent storage and manipulation of light with matter, remarkable advances have been made possible by EIT, which relies on quantum interference ~\cite{Lukin:2003ct,Fleischhauer:2005da}. The ATS regime emerges when this quantum interference is washed out due to the strong electromagnetic fields that induce a spectrally broad transparency window via the ac Stark effect. The crossover between EIT and ATS, identified with narrow and wide transparency, respectively, is an active topic of research~\cite{AbiSalloum:2010eg,Anisimov:2011fnb,Giner:2013,Tan:2014hk,Peng:2014gb,Lu:2015dk,He:2015ho}, begging the question of whether it is possible to directly leverage ATS for coherent control of light beyond treatment as an unfavorable EIT regime. Apart from a  theoretical proposal for a photon-echo interaction in the ATS regime~\cite{Liao:2014jr}, this possibility remains unexplored.

Here, we develop an approach where absorption of light pulses by dynamically controlled ATS lines mediates coherent storage, which is intrinsically suitable for efficient, broadband, and long-lived quantum memories. We demonstrate an experimental implementation of our protocol in a $\Lambda$-type three-level system of cold Rb atoms by reversible mapping of coherence from nanoseconds-long optical pulses on to a collective spin-state of the atoms for up to a microsecond of storage. Compared to established techniques~\cite{Liu:2001,Hosseini:2009, Afzelius:2010a, Reim:2010, Hedges:2010,Clausen:2011,Saglamyurek:2011}, our scheme is situated in a favorable regime for practical implementations, offering significantly relaxed requirements in terms of optical density, the power of the coupling electromagnetic field, technical complexity, robustness to decoherence, and experimental instabilities. Furthermore, we extend the approach to coherent manipulations of optical pulses, demonstrating spectral-temporal pulse shaping, temporal beam splitting, and interferometry. As our approach relies on a generic ATS (without additional control mechanisms or fragile quantum effects as required in EIT), it is readily achievable in diverse platforms including atomic and molecular vapours~\cite{Fleischhauer:2005da}, Rydberg atoms~\cite{Mohapatra:2007im,Pritchard:2010im}, optomechanical systems~\cite{Agarwal:2010dp,Huang:2010gv,Teufel:2011ih,SafaviNaeini:2011dm}, and superconducting quantum circuits~\cite{Sillanpaa:2009db,Abdumalikov:2010gw,Novikov:2013jt,Sun:2014kh}. These general principles can be further exploited for high-precision spectroscopy, metrology and  quantum technologies including quantum communication and optical quantum computing.

\section*{RESULTS}
\subsection*{The ATS quantum memory protocol}
\begin{figure*} [t!]
\begin{center}
\includegraphics[]{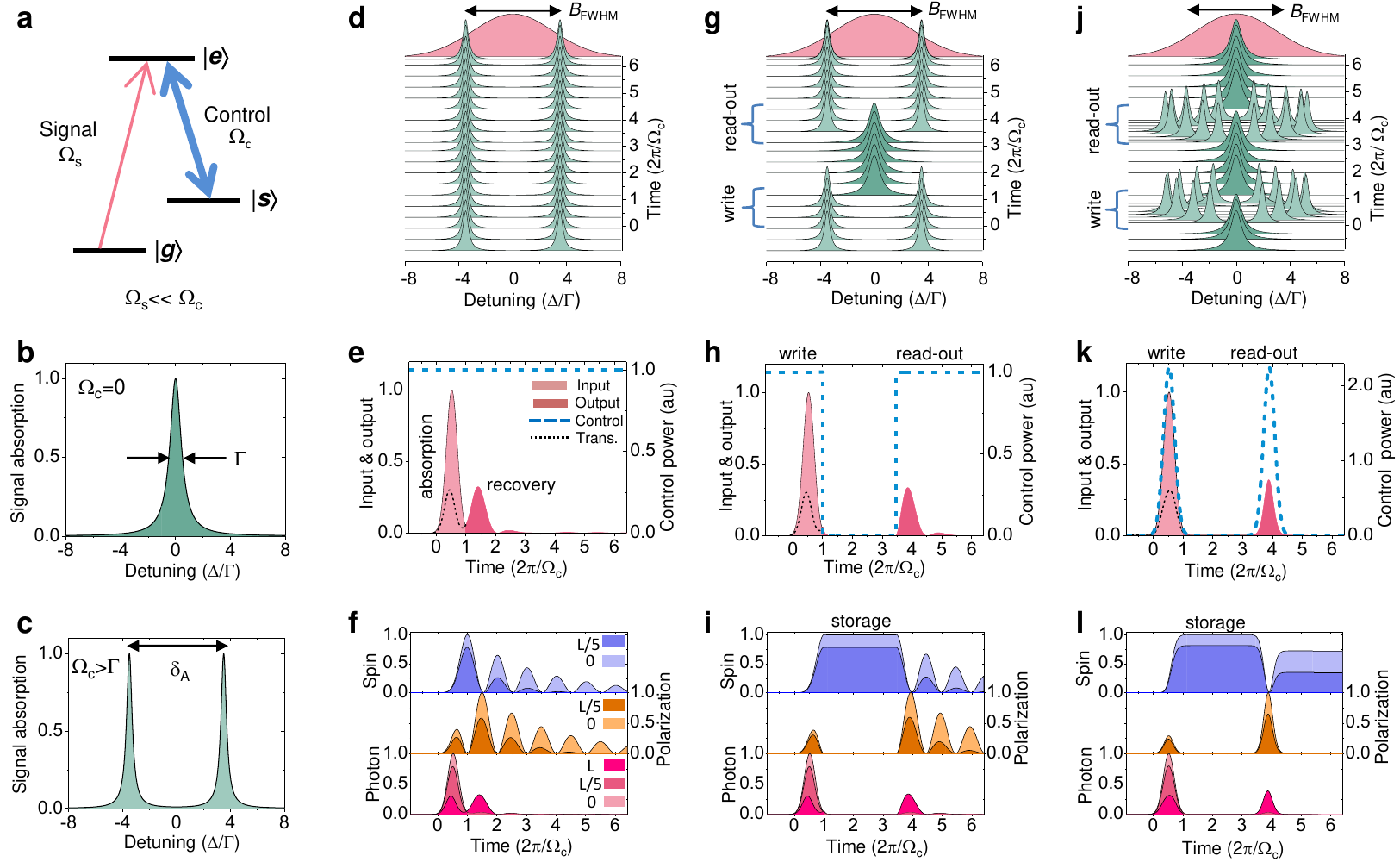}

\caption{\textbf{Autler-Townes splitting quantum memory protocol.}~\textbf{a},~ In a $\Lambda$-type three-level system, a strong coupling field (control, blue) and and weak probe field (signal, red) with Rabi frequencies $\Omega_{\rm c}$ and $\Omega_{\rm s}$ drive the transitions  $\ket{s}\leftrightarrow\ket{e}$ and $\ket{g}\leftrightarrow\ket{e}$, respectively.~\textbf{b}, When $\Omega_{\rm c}=0$, the absorption on $\ket{g}\leftrightarrow\ket{e}$ is naturally broadened with a linewidth of $\Gamma$.~\textbf{c},~When $\Omega_{\rm c}>\Gamma$, the natural absorption line is split into two with the peak spacing of $\delta_{\rm A}\approx\Omega_{c}$ and linewidths $\Gamma/2$ via the Autler-Townes effect.  Here, $\Omega_{\rm c}=7\Gamma$.~\textbf{d, e,}~A control field (blue dashed line in \textbf{e}), on at all times, generates a constant ATS. An input signal pulse with bandwidth of $B_{\rm FWHM}=\Omega_{\rm c}/2\pi$ (pink curve in \textbf{d}) is absorbed and repetitively recovered as output signals with delays of $2\pi/\Omega_{\rm c}$ as shown here for $\Omega_{\rm c}=7\Gamma$, $d=13$, and $\gamma_{\rm s}=0$. \textbf{f}, The system dynamics exhibit an oscillatory behaviour, as seen in the evolution of spin ($|S(z,t)|^{2}$), polarization ($|P(z,t)|^{2}$) and photonic ($|E(z,t)|^{2}$) coherences, near the entrance (light) and exit (dark) of the atomic medium.~\textbf{g-i}, Storage and on-demand recall of the signal is accomplished with an interrupted control, by switching the control off just before the first recovery (write), and switching it back on at a desired time (read-out).~\textbf{j-l}, Write and read-out can be achieved via pulsed control fields. Here, we illustrate this using control fields with the same gaussian profile as the signal, and with peak $\Omega_{\rm c}^{\rm peak}=10.5\Gamma$ to ensure the area $A_{\rm c}(\tau)= 2\pi$.}
\label{ATS} 
\end{center}
\end{figure*}

\noindent We begin with a simplified description of our approach for broadband light storage, with further details in Methods. We consider a three-level system in a $\Lambda$-configuration involving two long-lived spin levels $\ket{g}$ and $\ket{s}$ in the ground state and an excited state $\ket{e}$ that can be optically coupled to both ground levels, as shown in Fig.~\ref{ATS}a  (In principle, it is possible to implement this protocol in any three-level configuration with modifications.) Assuming that atoms initially populate  $\ket{g}$, a weak resonant signal field with Rabi frequency $\Omega_{\rm s}$  couples $\ket{g}$ to $\ket{e}$, and a strong resonant control field drives the transition between $\ket{s}$ and $\ket{e}$ with Rabi frequency $\Omega_{\rm c}$. The coherence decay rate for the optical transition $\ket{e}\leftrightarrow\ket{g}$ is $\gamma_{\rm e}={\Gamma}/{2}$, where $\Gamma$ is the radiative decay rate of state $\ket{e}$, and the rate for the spin transition $\ket{g}\leftrightarrow\ket{s}$ is $\gamma_{s}$. Provided that $\gamma_{\rm s}\ll\gamma_{\rm e}$, and that the control field has a Rabi frequency that is ideally much larger than the coherence decay rate of the optical transition ($\Omega_{\rm c}\gg\gamma_{\rm e}$), the $\ket{g}\leftrightarrow\ket{e}$ transition exhibits an Autler-Townes splitting whose peaks are separated by $\delta_{\rm A}\approx\Omega_{\rm c}$ (see Fig.~\ref{ATS}b and c). Note that the amount of ATS  can be dynamically changed using the control field power such that $\Omega_{\rm c}(t)\approx\delta_{\rm A}(t)=\alpha\sqrt{\mathcal{P}(t)}$, where $\alpha$ is a system-specific proportionality constant and $\mathcal{P}(t)$  is the control power. This feature plays an important role in the operation of our memory scheme, as detailed later.

For the ATS regime described above, we analyze the dynamics of the resonant atom-light system for $N$ uniformly distributed atoms using the Maxwell-Bloch equations~\cite{Gorshkov:2007gd,Liao:2014jr} :
\begin{align} \label{MB}
&(\partial_t + c \partial_z) \hat{E}(z,t) =  i g \sqrt{N} \hat{P}(z,t),\\
&\partial_t \hat{P}(z,t) = - \gamma_{e} \hat{P}(z,t) \!+\! i g \sqrt{N} \hat{E}(z,t) + \frac{i}{2} \Omega_{\rm c} \hat{S}(z,t),\\
\label{MB3}
&\partial_t \hat{S}(z,t) = -\gamma_\textrm{\rm s} \hat{S}(z,t)+ \frac{i}{2} \Omega_{\rm c}^* \hat{ P}(z,t),
\end{align}
where $\hat{E}(z,t)$ is the electric field operator for the photonic field, and $\hat{P}(z,t)$  and $\hat{S}(z,t)$ are the polarization and spin-wave operators, which are described by the collective atomic coherences of $\ket{g}\leftrightarrow\ket{e}$ and $\ket{g}\leftrightarrow\ket{s}$, respectively~\cite{Gorshkov:2007gd}. The strength of the atom-light coupling is $g\sqrt{N}=\sqrt{{c d  \gamma_{e}}/{2L}}$, where $d$ is the peak optical depth and  $L$ is the length of the atomic medium along $z$, which is the propagation direction of the photonic field. In the context of photon storage and recall, Equations~\ref{MB}-\ref{MB3} describe mapping coherence from an input (signal) photonic mode $\hat E_{\rm in}(z,t)=\hat E(0,t)$ (with a slowly varying temporal envelope over period $[0,\tau]$ such that~$\partial_z \hat{E}(z,t)\approx0$) to a spin-wave mode $\hat S(z,t>\tau)$, and after a storage time $T$, back to an output photonic mode $\hat E_{\rm out}(z,t)=\hat E(L,t>T)$. This reversible transfer between the photonic and spin components of the system is mediated by the evolution of the polarization $\hat P(z,t)$ and controlled by $\Omega_{\rm c}(z,t)$, which together with  $g\sqrt{N}$, $\gamma_{\rm e}$ and $\gamma_{\rm s}$, determine the output photonic mode. We note that the results of this treatment are applicable to both weak classical photonic fields (provided that $\Omega_{\rm s}\ll\Omega_{\rm c}$ and the photon number is much smaller than $N$) and quantum fields, including single-photon excitations.  

\

\begin{figure*}
\begin{center}
\includegraphics{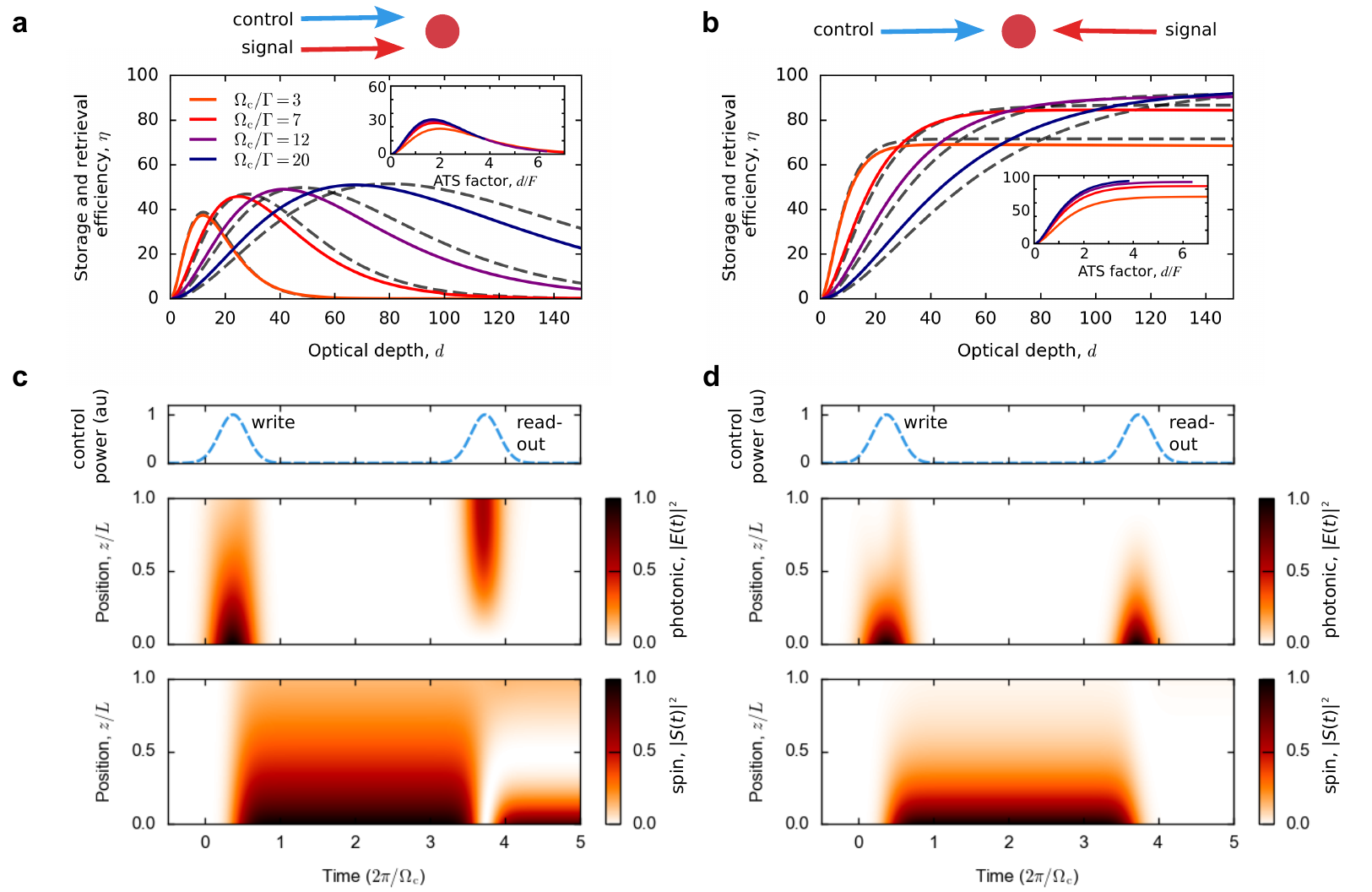}
\label{fig:efficiency}
\caption{{\bf Theoretical efficiency for storage and retrieval from the ATS memory.} {\bf a,b}, Numerically calculated efficiency vs.\ optical depth for co-propagating and counter-propagating signal--control beams is shown for different $F = \Omega_{\rm c}/\Gamma$ (solid curves), compared to calculations from simple model $\eta_{\rm f}=(d/2F)^2e^{-d/2F}e^{-1/F}$ and  $\eta_{\rm b} = (1-e^{-d/2F})^2e^{-1/F}$ (corresponding dashed curves), respectively. Each inset plots shows the same calculations with respect to effective optical depth, $\tilde{d} = d/2F$. The calculations are performed for a pulsed ATS memory with $A_{\rm c}^{\rm write}= A_{\rm c}^{\rm read}= 2\pi$, as in the main text. For the definition of $F=\Omega_{c}/\Gamma$ with this dynamic case, $\Omega_{\rm c}$ is effectively equivalent to a constant Rabi frequency that provides the same pulse area ($2\pi$). {\bf c, d},  Calculated temporal evolution of spin-wave $|S(z,t)|^{2}$ and photonic $|E(z,t)|^{2}$ modes during storage and retrieval is shown for the co-propagating and counter-propagating signal and control beam cases for all positions in the atomic medium (from $z=0$ to $L$), respectively. The calculations are performed for $d=40$, $F=12$ for the forward propagation mode, and $d=85$, $F=12$ for the backward propoagating mode, yielding $\eta_{\rm f}\approx 0.5$ and $\eta_{\rm b}\approx 0.9$, respectively. After the retrieval of the forward propagating mode, a significiant portion of the initial coherence remains in a spin-wave mode due to the re-absorption of the recalled signal near the entrance. For the backward retrieval, the transfer from the spin-wave mode back to photonic mode is  nearly perfect due to the time-reversal symmetry between the reading and writing that eliminates the re-absorption \cite{Gorshkov:2007gd}.}
\label{fig:efficiency}
\end{center}
\end{figure*}

With the goal of efficient light storage and retrieval under optimal conditions (i.e., moderate optical depths and control intensities), we investigate three related timing configurations for the input signal and control fields. In the first configuration (Fig.~\ref{ATS}d-f), a control field with constant power (constant $\Omega_{\rm c})$ generates a fixed ATS for all times. At $t=0$, a  short gaussian signal pulse whose bandwidth spans the ATS enters the medium [defining the input photonic mode $E(0,t)$]. In our protocol, we specify the following relationship between the signal bandwidth and ATS (Fig.~\ref{ATS}d): $\delta_{\rm A}/2\pi =\Omega_{\rm c}/2\pi= B_{\rm FWHM}$, where $B_{\rm FWHM}$ is the characteristic signal bandwidth at full-width half-maximum (FWHM), and $\tau_{\rm FWHM}=0.44/B_{\rm FWHM}$ is the corresponding time of the signal. As the signal propagates through the medium,  it is partially or fully absorbed by the ATS peaks, and the coherence carried by the signal pulse is mapped onto the spin levels over the total duration of the pulse $\tau\approx 2.25\tau_{\rm FWHM}$. Immediately after this mapping is complete, the signal pulse is coherently recovered between $t=\tau$ and $t=2\tau$, as shown in Fig.~\ref{ATS}e. An essential feature in this protocol arises from the spectral matching between ATS and the signal bandwidth ($2\pi {B_{\rm FWHM}}/{\Omega_{c}}= 1$): the pulse area for the control field $A_{\rm c}(t) = \int_0^{t} \Omega_{\rm c}(t^\prime) d t^\prime$ during the absorption and recovery, each occurring within $\tau$, is 
\begin{align}
A_{\rm c}(\tau)
&=\Omega_{c} \left({2.25}~{\tau_{\rm FWHM}}\right)=\Omega_{c}\left[2.25\left(\frac{0.44}{B_{\rm FWHM}}\right)\right] \approx 2\pi.
\label{area}
\end{align}
Complete analysis of the the atom-light system dynamics (illustrated in Fig.~\ref{ATS}f) show that the storage (absorption) and recovery processes rely upon the time-dependent oscillatory exchange of the coherence between the spin component and photonic components with a period $2\pi/\Omega_{\rm c}$, as described by the equations~\cite{Liao:2014jr} 
\begin{align}
\left|S(z,t)\right|^{2}\propto \cos^2{\left[{A_{\rm c}(t)}/{2}\right]}\label{eq:St}\\ 
\left|E(z,t)\right|^{2}\propto\sin^2{\left[{A_{\rm c}(t)}/{2}\right]}
\label{eq:Et}. 
\end{align}

 \begin{figure*} [ht!]
\begin{center}
\includegraphics[]{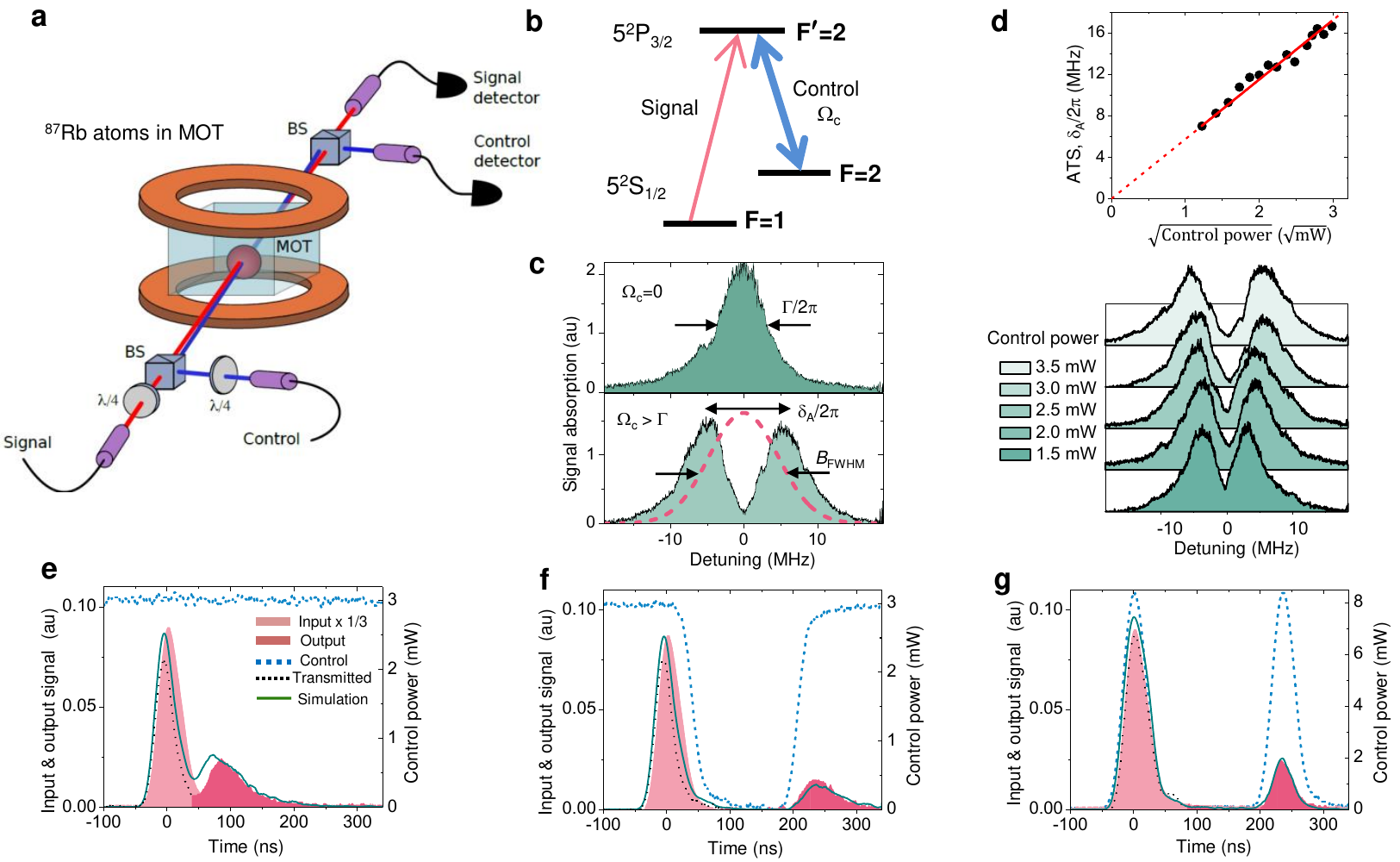}
\caption{\textbf{Experimental demonstration of ATS memory in cold atoms.}~\textbf{a,b}, A three-level $\Lambda$-system is formed using the D2 transition of $^{87}$Rb atoms that are cooled down in a magneto-optical-trap (MOT), as detailed in Methods. For storing light or probing ATS features in cold atoms, the linearly polarized  signal (red) and control (blue) beams from separate fiber-optics are combined on a free-space beamsplitter (BS), following quarter-waveplates ($\lambda/4$) for polarization control. Next, the signal and control beams, with separation angle 2$^{\circ}$, overlap inside the atomic cloud. After interaction with the atoms, the signal is coupled to a single-mode fiber for spatial filtering from the control field and directed to a photodiode for detection. A small fraction of the control beam reflected from the output beamsplitter is also detected for pulse synchronization and power calibration.~\textbf{c}, The transition linewidth ($\Gamma$) and a typical ATS ($\delta_{\rm A}$) for $F=1\leftrightarrow F^{\prime}=2$ are measured via the absorption of a weak frequency-swept signal, yielding $\Gamma^{\rm exp}/2\pi=(7.7\pm0.2)$~MHz for $\mathcal{P}=0$~mW (upper panel) and $\delta_{\rm A}/2\pi\approx\Omega_{\rm c}/2\pi=(10.7\pm 0.3)$~MHz for $\mathcal{P}=3.0$~mW (lower panel). \textbf{d}, The ATS is characterized with respect to control power, showing a good agreement with the linear relation $\delta_{\rm A}/2\pi=\alpha\sqrt{\mathcal{P}}$ where $\alpha$ is $5.75$~MHz/$\sqrt{\rm mW}$.~\textbf{e-g}, The storage and retrieval of an input signal pulse (bright shaded) with bandwidth  $B_{\rm FWHM}=11$~MHz~($\tau_{\rm FWHM}=40 $~ns), shown with the red dashed line in \textbf{c}, is demonstrated for the three timing configurations involving (\textbf{e}) constant, (\textbf{f}) interrupted and (\textbf{g}) pulsed  control fields (blue dashed), which are theoretically described in Fig.~\ref{ATS}e,h and k, respectively. The measurement results, including transmitted (black dotted) and recalled signal (dark shaded) are in agreement with the simulations of the Maxwell-Bloch equations (solid green) for experimental input and control traces, and $d^{\rm exp}=3.5$, $\Gamma^{\rm exp}/2\pi=7.7$, $\gamma_{\rm s}^{\rm exp}/2\pi=0.25$~MHz parameters, which were extracted from independent measurements} 
\label{setup}
\end{center}
\end{figure*}

\noindent As the coherence, initially carried by the input signal mode, reversibly evolves into the spin-wave and photonic modes inside the medium, it is mapped onto a delayed output signal mode (at $z=L$) for those times when the control pulse area $A_{\rm c}(t) = 2\pi n$ (where $n=1,2,3...$). Specifically, the initial coherence is transferred to the spin-wave mode in the first cycle ($n=1$, absorption), and it is converted back to a (output) photonic mode in the second  cycle ($n=2$, recovery) after a delay $T_{\rm delay} =2\pi/\Omega_{\rm c}\approx \tau$.
In general, some coherence remains as a spin-wave after the first recovery, and is periodically retrieved (with smaller intensity) as additional output signals at intervals of $T_{\rm delay}$.

In the second timing configuration, we convert this pre-determined delay process into an on-demand signal retrieval process with adjustable storage time. As illustrated in Fig.~\ref{ATS}g-i, this can be achieved  by  switching off the control field just before the recovery starts, and switching it back on after a desired time, which are referred to as ``write'' and ``read-out'' processes, respectively. The writing process transfers the optical coherence to the spin-wave mode [in the first $2\pi$-control-pulse area, $A_{\rm c}^{\rm write}(\tau)$] and traps it in the medium by decoupling the control. Then, after a storage time $T$, a second $2\pi$-control-pulse area [$A_{\rm c}^{\rm read}(\tau)$] initiates the reading process and  the stored coherence is released as the output photonic mode.

In the third timing configuration, we implement storage and retrieval of signal pulses with time-varying power [equivalently, $\Omega_{\rm c}(t)$] of the write and read-out fields, allowing pulsed operation of the ATS memory (Fig.~\ref{ATS}j-l). 
Since the dynamics depend only on $A_{\rm c}(\tau)$ (Eqs.~\ref{eq:St} and \ref{eq:Et}) and not the specific time-dependence of $\Omega_{\rm c}(t)$, the coherence will transfer from the photonic to spin mode, and back, 
each time
$A_{\rm c}^{\rm write}(\tau) = A_{\rm c}^{\rm read}(\tau) = 2\pi$. For example, the write and readout processes can be accomplished by using control pulses with the same temporal profile as the input signal, as shown in Fig.~\ref{ATS}k.  The pulsed operation of the ATS memory can be exploited for quantum signal processing, including beam-splitting and pulse shaping, as demonstrated in the next sections.

\begin{figure}[t!]
\begin{center}
\includegraphics{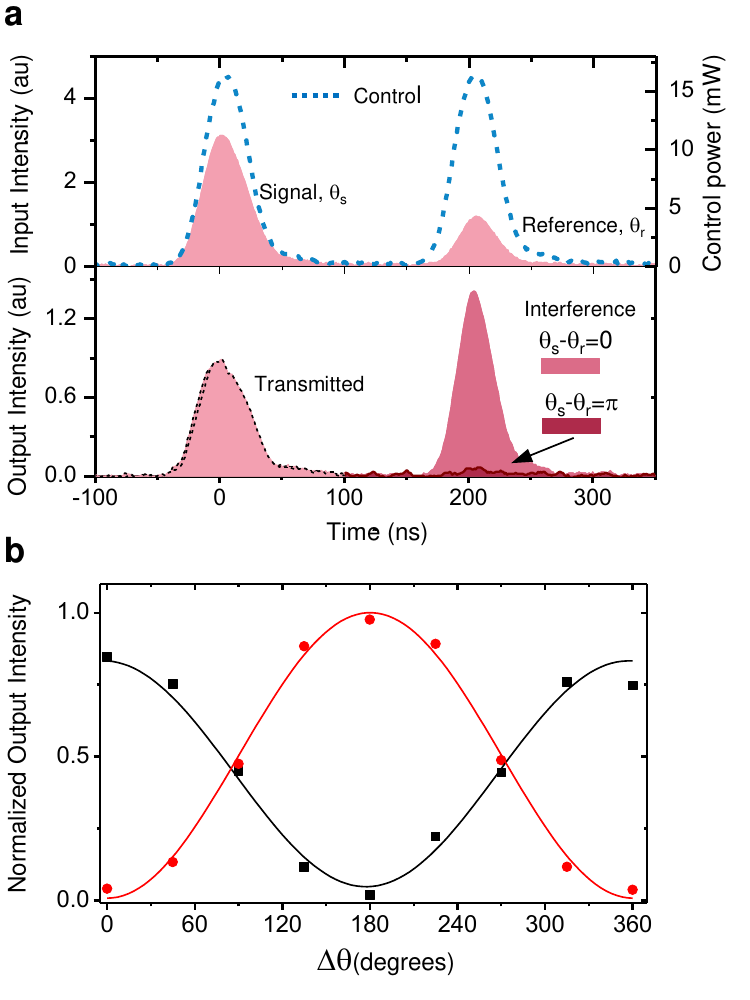}
\label{interference}
\end{center}
\caption{\textbf{Preservation of coherence in storage and recall}. \textbf{a}, A signal pulse, prepared with phase $\theta_{\rm s}$ is stored and retrieved after 200 ns using write and read-out control pulses, each encoded with the same phase. The phase of the retrieved signal $\theta_{s}^{\prime}$ is determined by three contributions: the phase of the input signal, the  accumulated phase during the evolution of spin-wave $\theta_{\rm sw}$ for a given storage time, and the phases of the write and read-out $\theta_{\rm c}$, yielding $\theta_{\rm s}^{\prime}=\theta_{\rm s}+\delta\theta$, where $\delta{\theta}=\theta_{\rm sw}+\theta_{\rm c}$ is the overall phase imparted in storage and retrieval, which is controllable via $\theta_{\rm c}$. At the moment of recall, a reference pulse with the same duration as the signal, but with a fixed phase $\theta_{\rm r}$  is sent to the medium, as shown in the upper panel. Thus, the transmitted portion of the reference pulse (adjusted to have the same amplitude of the retrieved signal) and recalled signal pulse interfere. The intensity of the interfering light pulses depends on only $\theta_{\rm s}$ with respect to $\theta_{\rm r}$, if the memory introduces the same $\delta{\theta}$ unitary phase transformation for every $\theta_{\rm s}$ with a fixed $\theta_{\rm c}$ and storage time. Two such interference measurements are illustrated in the lower panel for $\Delta\theta=\theta_{\rm s}-\theta_{\rm r}=0$ and $\Delta\theta=\pi$ with an observation of near-perfect constructive and destructive interference, respectively, showing that $\delta\theta=0$ for each $\theta_{\rm s}$, and thus the relative phase is preserved in the memory. \textbf{b}, $\theta_{\rm s}$ is varied from $0^{0}$ to $360^{0}$ in steps of $45^{0}$ and the resulting intensity measurements ($I$) yield an interference visibility $V=(I_{\rm max}-I_{\rm min})/(I_{\rm max}+I_{\rm min})=89\pm6~\%$ and $V=98\pm2~\%$ for $\delta\theta=0$ (black square) and $\delta\theta=\pi$ (red circle), determined by fitting each data set to a sinusoidal function, respectively.}
\label{interference}
\end{figure}

The total efficiency of the storage and recall processes can approach unity ($\eta = 1$) under appropriate conditions, as detailed in Methods. The first condition is to provide a sufficiently large optical depth and a sufficiently large and uniform Rabi frequency such that the pulse area for both the write and read-out control fields are near $2\pi$, and the ATS factor $F =\Omega_{\rm c}/\Gamma\gg1$. Note that because the memory's acceptance bandwidth is directly determined by $\Omega_{\rm c}$ ($B_{\rm FHWM}\approx\Omega_{\rm c}/2\pi$, which can be chosen at will), the requirement of large $F$ makes the ATS memory protocol intrinsically suitable for efficient and broadband operation. However, as shown Fig.~\ref{fig:efficiency}a and c, the maximum efficiency for the first-order recall ($n=2$) is limited to $54\%$ due to the re-absorption of retrieved light inside the medium, as in other resonant absorption-based memory protocols such as photon-echo based techniques~\cite{Afzelius:2009, Tittel:2010}. The second condition for large efficiency is to eliminate the re-absorption, which is possible, for example, by using  backward retrieval via counter-propagating write and read-out control fields \cite{Gorshkov:2007gd, Afzelius:2009}. In this arrangement, as shown in Fig.~\ref{fig:efficiency}b,d,  near-unity memory efficiency ($\eta\approx 0.9$) is possible with moderate optical depths ($d\rightarrow85$), which is technically feasible in several platforms \cite{Riedll:2012,Choo:2016}. Finally, the ATS memory provides optimal operating conditions in terms of technical requirements, including much broader bandwidth compared to the EIT scheme~\cite{Lukin:2001, Lvovsky:2009}, at the cost of control fields with larger strength. At the same time, demands on the strength of the control field and optical depth are significantly reduced compared to off-resonant Raman memory schemes that can have the same bandwidth~\cite{Lvovsky:2009, Reim:2010}.

\subsection*{Experimental setup and ATS characterization} 

\noindent We demonstrate a proof-of principle of the ATS memory protocol using an ensemble of $\sim2.5\times10^8$ cold $^{87}$Rb atoms released from a magneto-optical trap (MOT). Fig.~\ref{setup}a,b shows a simplified diagram of our experimental setup, detailed in Methods. In this system, we construct a $\Lambda$-configuration using two ground hyperfine levels ($F=1,2$ as $\ket{g}$ and $\ket{s}$ with spacing of 6.83~GHz) and one excited  ($F^{\prime}=2$ as $\ket{e}$) level of the $^{87}$Rb D2 transition ($780$~nm). Nearly co-propagating signal and control fields (derived from continuous-wave lasers whose amplitudes, frequencies and phases are controlled via acousto-optic modulators) are coupled to the  $\ket{F=1}\rightarrow 
\ket{F^{\prime}=2}$ and $\ket{F=2}\rightarrow 
\ket{F^{\prime}=2}$ transitions, respectively. Absorption measurements of the  $\ket{F=1}\rightarrow 
\ket{F^{\prime}=2}$ transition without any control field yield a single absorption feature  with a peak optical depth of $d^{\rm exp}=3$ to 3.5 and linewidth of  $\Gamma^{\rm exp}/2\pi = (7.7\pm0.2)$~MHz, including a residual Doppler broadening (upper panel in Fig.~\ref{setup}c). We measure the same transition in the presence of a strong control field and observe two well-separated absorption peaks, the manifestation of ATS. Our typical control power $\mathcal{P}=3.0$~mW gives a peak separation $\delta_{\rm A}/2\pi = (10.7\pm0.3)$~MHz (lower panel in Fig.~\ref{setup}c), showing that our system is predominantly in  the ATS regime: $\Omega_{\rm c} > \Gamma^{\rm exp}$~\cite{Anisimov:2011fnb,Giner:2013}. Note that the ATS memory protocol still works in the regimes of $\Omega_{\rm c}\approx \Gamma$ (the ATS-EIT crossover regime) and  $\Omega_{\rm c} < \Gamma$ (the EIT regime), but with lower efficiencies as discussed in Methods. Further, we characterize the peak separation $\delta_{\rm A}$ with respect to the control power as shown in Fig.~\ref{setup}d.

\begin{figure*}
\includegraphics{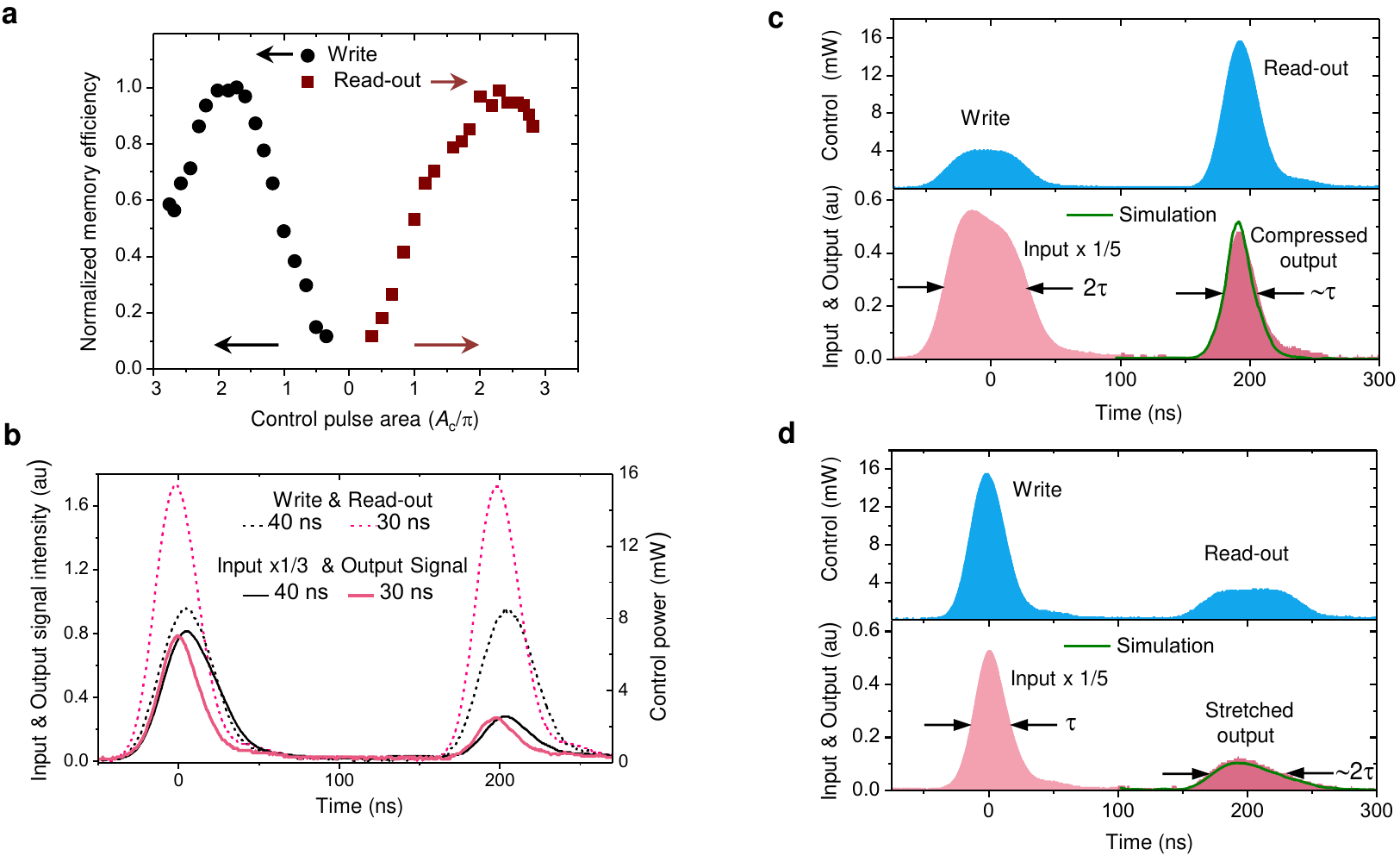}
\caption{\textbf{Dynamic control of memory bandwidth and temporal shaping of signal pulses} \textbf{a}, The memory efficiency with respect to  write pulse area ($A_{\rm c}^{\rm write}$) and read-out pulse area ($A_{\rm c}^{\rm read}$), each with the same profile as the input signal, is independently characterized for storage of a gaussian signal pulse with $\tau_{\rm FWHM}=40$ ns. For each pulse-area setting, the required peak Rabi frequency and peak power are calculated from $A_{\rm c}(\tau)=\Omega_{c}^{\rm peak}\sqrt{\pi/(2\ln{2})}\tau_{\rm FWHM}$ and $P_{c}^{\rm peak}=(\Omega_{c}^{\rm peak}/\alpha)^2$ where $\alpha/2\pi=\rm 5.75~ \rm MHz/\sqrt{\rm mW}$ (Fig.~\ref{setup}c). \textbf{b}, The dynamic control of the acceptance bandwidth is demonstrated for 11.0 MHz and 14.7 MHz input signals ($\tau_{\rm FWHM}=40$ ns and $\tau_{\rm FWHM}=30$ ns) using write pulses with profiles matching the input pulses and the peak powers of 8.5~mW and 15.0~mW, respectively, such that $A_{c}^{\rm write}=2\pi$. After 200 ns storage, a read-out pulse with the same profile as the write pulse is applied, leading to recalled signals with efficiencies of $8.8\%$  and $8.3\%$ for the memory bandwidths of 11 MHz and 14.7 MHz, respectively. \textbf{c,d}, Temporal compression (stretching) of an input signal pulse with  $\tau_{\rm FWHM}=30$~ns ($\tau_{\rm FWHM}=60$~ns) is demonstrated in \textbf{c} (\textbf{d}). This is accomplished with optimal efficiency by storing the signal with a write pulse of the same duration as the input and  $A_{\rm c}^{\rm write}=2\pi$, but retrieving it using a read-out pulse of shorter (longer) duration than the input while maintaining $A_{\rm c}^{\rm read}=2\pi$. The resulting output is a  compressed (stretched) pulse with  duration of  $\tau_{\rm FWHM}=28$~ns ($\tau_{\rm FWHM}=58$~ns) and efficiency of  $7.4\%$ ($8.6\%$), yielding about factor of 2 compression (stretching). The experimental results are well simulated using the parameters of $d^{\rm exp}=3.5$, $\Gamma^{\rm exp}/2\pi=7.7$~MHz and $\gamma_{\rm s}^{\rm exp}/2\pi=0.35$~MHz, as shown in solid green lines}
\label{bandwidth}
\end{figure*}

\subsection*{Demonstration of ATS memory in cold atoms}
 We implement the ATS memory using the three timing configurations described above. First, we examine the absorption and recovery of a short gaussian pulse with a constant control power ($\mathcal{P} = 3.0$~mW) which generates a static ATS with $\delta_{\rm A}/2\pi\approx11$~MHz, as shown in Fig.~\ref{setup}c. We set the signal bandwidth  $B_{\rm FWHM} = 11$~MHz ($\tau_{\rm FWHM} = 40$~ns)  to fulfill the bandwidth-matching condition : $B_{\rm FWHM}\approx\delta_{\rm A}/2\pi =\Omega_{\rm c}/2\pi$ (red dashed trace in Fig.~\ref{setup}c). We observe that upon entering the medium, the pulse is partially absorbed and then is emitted with $\eta^{\rm exp}= 13.0\%$ after a delay of one pulse duration (Fig.~\ref{setup}e), showing good agreement with our prediction and simulations based on the numerical analysis of the Maxwell-Bloch equations~\cite{SI}.
 
 To demonstrate the second timing configuration with recall on-demand, we repeat the above procedure but switch off the control field just before the recovery starts, trapping the coherence in the spin-wave mode. After a storage time $T= 230$~ns,  we switch the control field back on, resulting in on-demand retrieval with $\eta^{\rm exp}=7.3\%$, as shown in Fig.~\ref{setup}f. 
 
 To demonstrate the third timing configuration corresponding to pulsed operation, we generate a write control field with the same temporal profile and duration as the input signal ($\tau_{\rm FWHM} =  40$~ns) and overlap it with the input signal inside the medium, which leads to storage. After $T = 230$~ns, we recall this signal by sending a read-out control pulse with the same gaussian shape and duration as the write pulse, as shown in Fig.~\ref{setup}g. In order to maintain the $A_{\rm c}^{\rm write} =A_{\rm c}^{\rm read} = 2\pi$ pulse area, we set the peak power for the read and write pulses to $\mathcal{P}_{\rm peak} = 8.5$ mW (corresponding $\Omega_{\rm c}/2\pi = 17$~MHz). This gives nearly the same optimal recall efficiency ($\eta^{\rm exp}=7.8~\%$) as obtained in the second  configuration.
 
 The main limiting factors to our low memory efficiency are large spin-wave decoherence rates ($\gamma_{s}^{\rm exp}/2\pi=0.24$ to 0.40~MHz) and small optical depths in our current system. These factors also limit the observed storage times up to a microsecond~\cite{SI}.
However, significant improvement for our memory performance is readily achievable, as detailed in Methods and \cite{SI}.

\begin{figure}[]
\begin{center}
\includegraphics{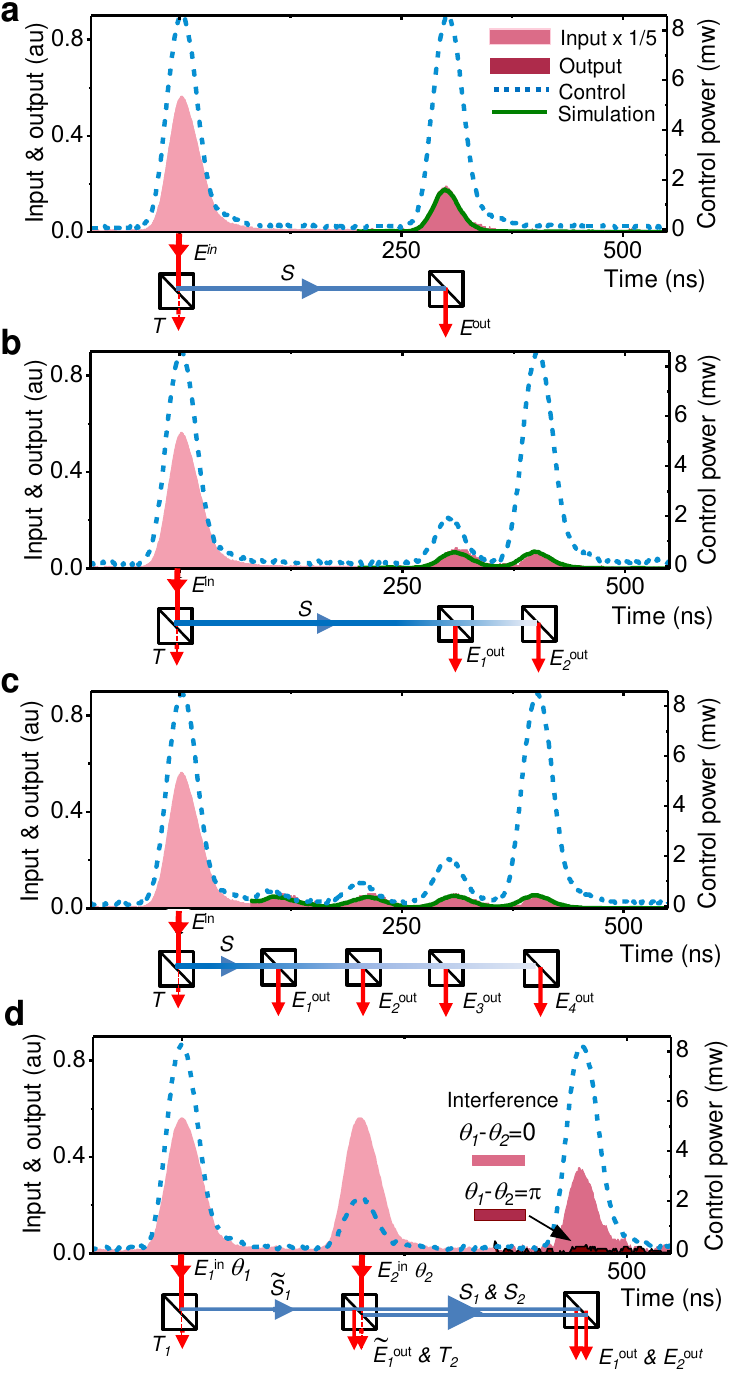}

\label{splitter}
\end{center}
\caption{\textbf{Demonstration of temporal beam splitting.} \textbf{a, b, c,} Initial coherence from an input photonic mode $E^{\rm in}$ is partially mapped onto a spin-wave mode $S$ and transmitted photonic mode $T$ (due to non-unity absorption probability in our memory) using a write pulse with the area of $2\pi$. The stored coherence in $S$ is transformed back to a single ($E^{\rm out}$), two ($E_{1}^{\rm out}$ and $E_{2}^{\rm out}$) or four ($E_{1}^{\rm out}$, $E_{2}^{\rm out}$, $E_{3}^{\rm out}$, $E_{4}^{\rm out}$) output photonic modes using one, two or four read-out pulses with appropriately selected areas ($\leq2\pi$) as shown in \textbf{a, b} and \textbf{c}, respectively. The amplitude of the  output signals are adjusted to be nearly the same by selecting these pulse areas and considering the spin-wave decay time, but they can be arbitrarily changed. The experimental results are well simulated using the parameters $d^{\rm exp}=3.5$,  $\gamma_{s}^{\rm exp}/2\pi=0.32$~MHz and $\Gamma^{\rm exp}/2\pi=$7.7 MHz as shown with green solid lines. \textbf{d,} Coherence from two distinct input photonic modes $E_{1}^{\rm in}$ and $E_{2}^{\rm in}$ with phases of $\theta_{1}$ and $\theta_{2}$, is mapped onto two spin-wave modes $\tilde{S_{1}}$ and $S_{2}$ using the first and second control pulses with the same phase and the areas of $2\pi$ and $\pi$, respectively. After the second control, the resulting spin-waves $S_{1}$ and $S_{2}$ interfere with equal amplitudes. The third control pulse transforms coherence from the interfering spin-waves to an output photonic mode whose intensity depends on ($\theta_{1}$- $\theta_{2}$) as shown for $\Delta\theta=\theta_{1}-\theta_{2}=0$ and $\Delta\theta=\pi$}
\label{splitter}
\end{figure}

\subsection*{Preservation of coherence after storage and recall}
\noindent Next, we experimentally verify the phase-preserving nature of the ATS memory. To do so, we store and recall $\tau_{\rm FWHM} =40$ ns signal pulses with different phase settings. For each signal pulse, we send a reference pulse at the recall time that has the same duration but a fixed phase. In this way, the recalled signal temporally overlaps and interferes with the transmitted portion of the reference pulse, as shown in Fig.~\ref{interference}a. We observe near-perfect constructive and destructive interference for phase differences of zero and $\pi$, respectively, and measure an average interference visibility of $94\%$, as illustrated in Fig.~\ref{interference}b.

\subsection*{Dynamically controllable storage bandwidth}

The acceptance bandwidth of an ATS memory can be dynamically controlled. As described above, optimal writing requires a control area $A_{\rm c}^{\rm write}(\tau) = \int_0^{\tau} \Omega_{\rm c}(t) d t=2\pi$,  where $\tau\propto1/B_{\rm FWHM}$ is the input signal duration that temporally overlaps with the write field. Increasing the bandwidth of an ATS memory (decreasing $\tau$) requires increasing $\Omega_{\rm c}(t)\propto \sqrt{\mathcal{P}}$. To demonstrate this feature, we first measure the memory efficiency as a function of the write pulse area, verifying that optimal pulse area for the write is $2\pi$ (Fig.~\ref{bandwidth}a). Next, we increase the bandwidth of the input signal from  $B_{\rm FWHM} = 11$~MHz ($\tau_{\rm FWHM} = 40$~ns) to $B_{\rm FWHM}\approx 15$~MHz ($\tau_{\rm FWHM} = 30$~ns). Accordingly, we decrease the duration of the write pulse to 30 ns (at FWHM), and increase its peak power from $\mathcal{P}_{\rm peak} = 8.5$~mW  to $15$~mW such that the area is maintained at $2\pi$. After 200 ns storage, retrieval is achieved via a read-out pulse with the same profile as the write, resulting in the optimal storage efficiency of $8.4\%$. 

\subsection*{Temporal compression and stretching}

The pulse-area-based operation of an ATS memory can be used for manipulating the temporal profile of optical pulses. As in the writing process,  optimal read-out requires the pulse area  $ A_{\rm c}^{\rm read}(\tau)=2\pi$, which we experimentally confirm by measuring  the storage efficiency versus read-out pulse area (Fig.~\ref{bandwidth}a). In contrast, the read-out pulse duration does not need to match the input signal's duration. Provided that $\Omega_{\rm c}(t)$ is set to give $A_{\rm c}^{\rm read}(\tau) = 2\pi$, the read-out can be longer or shorter than the input, leading  to temporally stretched or compressed output signals, respectively. To demonstrate this, we store a $\tau_{\rm FWHM}=60$~ns (or $\tau_{\rm FWHM}=30$~ns) signal pulse using a 60 ns (or 30 ns) write pulse, but retrieve it using a 30 ns (or $\approx60$~ns) read-out pulse, leading a temporally compressed (or stretched) output by a compression (stretching) factor of 2, as shown in Fig.~\ref{bandwidth}c and d, respectively. This pulse-shaping capability may be used for bandwidth matching between different photonic and atomic systems \cite{Hosseini:2009, Saglamyurek:2014}.

\subsection*{Temporal beam splitting}

Finally, we demonstrate that an ATS memory can serve as a network of  reconfigurable $2\times2$ temporal beam splitters that have arbitrary splitting ratios and phase control. In this scheme, the spin-wave mode and photonic mode correspond to the input and output ports of the ATS temporal beam splitter. Each control pulse (write or read-out) functions as a beam-splitter, performing a unitary transformation of the coherence between the input and output ports. The pulse area, from zero to $2\pi$, determines the splitting ratio between the photonic mode (retrieved or transmitted) and the spin mode. In this way, the coherence of an input photonic mode can be mapped onto the spin-wave modes and output photonic modes with desired fractions using multiple control pulses (beam splitters), each having an appropriate pulse area, as experimentally demonstrated in Fig.~\ref{splitter}. In these measurements, the initial coherence is recalled in one, two or four distinct temporal modes with nearly equal amplitudes, as illustrated in Fig.~\ref{splitter}a-c, respectively.

The beam splitting operation of the ATS memory can be used for interfering optical signals from distinct temporal modes, suitable for a time-domain version of Hong-Ou-Mandel-type interference or manipulation of time-bin photonic qubits. To demonstrate this capability (Fig.~\ref{splitter}d), two control fields with appropriately selected pulse areas are used to map coherence from two equal-amplitude time-separated input photonic modes ($E_{1}^{\rm in}$ and $E_{2}^{\rm in}$ with phases of $\theta_{1}$ and $\theta_{2}$, respectively) onto two equal-amplitude spin modes ($S_{1}$ and $S_{2}$) that interfere in the medium. The coherence from the interfering spin-waves is converted back into a single output photonic mode, whose amplitude depends on the relative phase of the input modes and control fields. This demonstrates the possibility of performing photonic interference using an ATS memory, which could find use in quantum communication and optical quantum computing \cite{Saglamyurek:2014, Reim:2013, Campbell:2014, Heshami:2016}.

\section*{DISCUSSION}

The ATS memory protocol combines favorable features of other memory techniques in a single implementation, relaxing the technical requirements for practical quantum memories. Similar to the EIT scheme \cite{Lukin:2001}, ATS memory is an on-resonant interaction scheme with large light-matter coupling, but it operates in a completely different regime. An EIT memory relies on a dispersion feature, leading to the slow-light-based signal transmission through a narrow transparency window, where $\Gamma\gg\Omega_{\rm c}\gg2\pi B_{\rm FWHM}$. In contrast, an ATS memory is based on signal absorption over a broad bandwidth, where $\Gamma\ll\Omega_{\rm c}\approx 2\pi B_{\rm FWHM}$. Since the ATS memory is not subject to slow-light requirements for storage,  it is much more robust against spin decoherence and laser frequency instabilities. Moreover, similar to the off-resonant Raman memory \cite{Reim:2010}, the ATS memory offers dynamically controllable large bandwidths. However, the resonant nature of the ATS scheme relaxes the requirements for  large optical depth and  control field power compared to the broadband Raman memory. Finally, the absorption-based operation and the efficiency scaling of an ATS memory exhibits the same character as the atomic frequency comb (AFC) protocol \cite{Afzelius:2010a}, which relies on the absorption of a signal through a comb-shaped spectrum. The absorption peaks in an AFC with rare-earth ions are obtained using advanced persistent hole burning techniques, whereas in the ATS protocol they are generated via the ac-Stark splitting, which is technically much less demanding and more versatile, allowing dynamic control of the peak spacing. In contrast to the AFC, the ATS approach is not a spectral or temporal multimode scheme, though this limitation can be circumvented via spatial multiplexing techniques \cite{Kuzmich,Pu:2017}. 

Finally, we experimentally evaluate the performance of the ATS protocol compared to EIT and off-resonant Raman approaches in our cold-atom system, featuring $\Gamma^{\rm exp}/2\pi = 7.7$~MHz, $\gamma_{\rm s}^{\rm exp}/2\pi \approx 0.4$~MHz and $d^{\rm \exp}\approx3.3$. First, we implement EIT-based slow-light by generating a transparency window as narrow as $2.5-3.0$ MHz at FWHM (limited by the spin decoherence and laser frequency instability) for $\mathcal{P}=0.70$~mW. This results in a group delay of $\approx70$~ns for a signal field with $B_{\rm FWHM}\approx1.2$~MHz bandwidth ($\tau_{\rm FWHM}=380$ ns). With these conditions, it is not possible to stop the signal for on-demand recall due to insufficient optical depth and a relatively large transparency window. Next, we implement the off-resonant Raman scheme for storage of a signal with our typical bandwidth $B_{\rm FWHM}=11$~MHz. To do so, we red-detune the signal and control fields by $22$~MHz, and  increase the peak power of control fields to $8.5$~mW. We retrieve the signal after $200$~ns storage with the same time-bandwidth product as the ATS memory, but with only $\approx2\%$ efficiency. This is significantly less than the typical efficiency of the ATS memory ($7~\%$) for the same storage time. These results show that even with limited resources and technical constraints, the ATS scheme offers a workable and robust light-matter interface for fundamental research and potential applications.

In conclusion, we have introduced and experimentally demonstrated a novel light storage and manipulation technique based on dynamically controlled Autler-Townes splitting. Our approach is inherently suitable for the realization of practical quantum optical devices in variety of atomic and molecular systems.  We anticipate that  our investigation will spur further fundamental studies and will bring new possibilities in precision spectroscopy, metrology, and quantum information processing.

\begin{center}
\textbf{METHODS}
\end{center}

\section*{ATS memory efficiency}
We define the total memory efficiency $\eta$ to be the ratio of the number of photons in the output mode [$E(L,t>T)$, where $T$ is the storage time] to the number of photons in the input mode [$E(0,t<\tau)$, where $\tau$ is the pulse duration]. As in the general case for spin-wave storage, this efficiency is determined by three factors: the storage efficiency $\eta_{\rm s}$, the ratio of the number of stored spin excitations to the number of incoming photons; the retrieval efficiency $\eta_{\rm r}$, the ratio of the number of retrieved photons to the number of stored spin excitations; and the spin-wave survival efficiency during storage $\eta_{\rm d}$, which is determined by decoherence effects that arise in practical settings.  Together, the overall efficiency is the product of these three
\begin{align}
\eta=\frac{\int_{T}^{\infty}|E(L,t)|^{2} dt}{\int_{0}^{\tau}|E(0,t)|^{2}dt}=\eta_{\rm s}\eta_{\rm r}\eta_{\rm d}.
\label{eff}
\end{align}
In the ATS memory, the reversible transfer of coherence between the photonic and spin-wave modes is mediated by the evolution of the coherence on to the collective state between $\ket{e}$ and $\ket{g}$, which describes the polarization mode. In the writing stage, the initial coherence from the input photonic mode is mapped on to the polarization (the absorption process) that simultaneously begins its evolution to the spin-wave mode (as per Eqs.~\ref{eq:St} and \ref{eq:Et}), and initiates storage. Thus the storage efficiency $\eta_{\rm s}$ depends on two factors: first, the absorption factor $\mu_{\rm abs}$ is the coherent sum of the absorption probabilities of the input photon by each atom in the ensemble (determined by $d$, $\Omega_{c}^{\rm write}$ and $\Gamma$); second, $\mu_{\rm w}$ is the efficiency of the polarization-mediated reversible  transfer between the photonic and the spin-wave modes [determined by $A_{\rm c}^{\rm write}(\Omega_{\rm c})$]. When $\mu_{\rm w}$ approaches unity, the number of excitations left in the polarization mode at the completion of the writing stage is minimal. 

The same treatment can be applied for the read-out stage, in which coherence stored in the spin-wave mode is transferred  to the polarization that simultaneously emits photons in the output mode (the re-emission process). In this case, the retrieval efficiency $\eta_{\rm r}$ depends on $\mu_{\rm re}$, the collective re-emission probability of the polarized atoms across the ensemble (determined by $d$, $\Omega_{\rm c}^{\rm read}$, $\Gamma$); and  $\mu_{\rm r}$, the efficiency of the polarization-mediated reversible transfer between the  spin-wave and photonic modes [determined by $A_{\rm c}^{\rm read}(\Omega_{\rm c}$), in the same fashion of $\mu_{\rm w}$.] 

Using the terms stated above and assuming that spin-wave decoherence is neglibly small ($\eta_{\rm d}=1$), Eq.~\ref{eff} is
\begin{align}
\eta=\mu_{\rm w}(A_{\rm c}^{\rm write}) \mu_{\rm abs}(d,\Gamma,\Omega_{\rm c}^{\rm write}) \mu_{\rm re}(d,\Gamma, \Omega_{\rm c}^{read}) \mu_{\rm r}(A_{\rm c}^{\rm read}).
\label{eff2}
\end{align}
In the particular case when the pulse area for both writing and read-out fields is $A_{\rm c}^{\rm write} = A_{\rm c}^{\rm read }= 2\pi$, no polarization remains in the medium right after the writing and reading, thereby achieving complete transfer of coherence between the photonic and  spin-wave mode. Thus, $\mu_{\rm w}$ and $\mu_{\rm r}$ approach unity, and the memory efficiency simplifies to
\begin{align}
\eta=\mu_{\rm abs}\mu_{\rm re},
\label{eff3}
\end{align}
which is solely determined by the collective absorption and re-emission processes. This description is common in resonant absorption-based quantum memory approaches, such  as the AFC and Controlled Reversible Inhomogeneous Broadening (CRIB) protocols that have been well analyzed~\cite{Afzelius:2009, Tittel:2010}. The analytic solutions describing the memory efficiency for the forward propagating mode ($\eta_{\rm f}$) and backward propagating mode ($\eta_{\rm b}$) are
\begin{align}
\eta_{\rm f}=\tilde{d}^2 e^{-\tilde{d}}\mu_{\rm d}\label{eq:effForward1}\\
\eta_{\rm b}=(1-e^{-\tilde{d}})^{2}\mu_{\rm d}
\label{eq:effBack1}
\end{align}
where $\tilde{\rm d}$  and $\mu_{\rm d}$ represent effective optical depth and collective coherence survival probability for polarization (during writing and reading), respectively. Since the absorption and re-emission occur via  ATS lines that span the signal spectrum $B_{\rm FWHM}$, $\tilde{d}$ is an increasing function of peak optical depth ($d$) and the linewidth of the ATS peaks ($\Gamma/2$), and a decreasing function of the spacing between the ATS peaks ($\delta_{\rm A}\approx\Omega_{\rm c}$). Note that for the dynamic ATS cases, the peak spacing is effectively equivalent to the peak spacing of the fixed ATS that provides the same pulse area. For $B_{\rm FWHM}=\Omega_{\rm c}/2\pi$ (equivalently $A_{\rm c}^{\rm write}=A_{\rm c}^{\rm read}=2\pi$), the effective optical depth is given by $\tilde{d}=d/2F$  where $F=\Omega_{\rm c}/\Gamma$ is the ATS factor. Decoherence (dephasing) of the polarization in the absorption and re-emission processes (which reduces $\mu_{d}$) occurs due to the finite width of the ATS lines. Hence, $\mu_{\rm d}$ is determined by the Fourier transform of an ATS line over the interaction time (input and output signal duration). This is equal to $e^{-1/F}$ for an ATS line, due to its Lorentzian lineshape. Using these definitions, the efficiency of the ATS memory under the specified conditions can be approximated as 
\begin{align}
\eta_{\rm f}\approx{(d/2F)}^2 e^{-d/2F} e^{-1/F} \label{eq:effForward2}\\
\eta_{\rm b}\approx(1-e^{-d/2F})^{2}e^{-1/F}.\label{eq:effBack2}
\end{align}

The efficiency results calculated from these expressions show reasonably good agreement with the numerically calculated efficiencies for a wide range of $d$ and $F$, as presented in the main text with Fig.~\ref{fig:efficiency}a,b. The inset of each figure illustrates the efficiency with respect to $\tilde{d}$, which provides additional insight together with Eq.~\ref{eq:effForward1} and \ref{eq:effBack1}. For the forward-propagating output mode and a given $F>1$, the efficiency of the ATS memory increases until  $\tilde{d}\approx1.5$ (the optimal effective optical depth) as seen in Fig.~\ref{fig:efficiency}a. In this regime, the first terms of Eqs.~\ref{eq:effForward1} and \ref{eq:effForward2} dominate because large optical depths (number of atoms) enhance the  collective absorption and re-emission probability. As the optical depth is further increased, the efficiency decreases since the second exponentially decaying term dominates. This term effectively describes the re-absorption probability of the emitted photons, which are converted to the spin-wave excitation and remain in the medium after read-out. For this reason, the efficiency for the forward-propagating mode is limited to $54\%$, which is achieved when  $\tilde{d}\approx1.5$ and $F$ is sufficiently large. In contrast, for the backward propagating mode, the re-absorption term is eliminated due to the fact that constructively interfering re-emission occurs only at the input side of the medium. Thus, the efficiency is an always-increasing function of $\tilde{d}$ (first factor in Eq.~\ref{eq:effBack1} and \ref{eq:effBack2}), and it tends to saturate for  $\tilde{d}\gtrsim 3$, as seen in Fig.~\ref{fig:efficiency}b. In addition to these factors, the efficiency strongly depends on the ATS factor $F$, which directly determines the polarization dephasing (the last term of Eqs.~\ref{eq:effForward2}  and  \ref{eq:effBack2}). In general for a given $\tilde{d}$, the total efficiency increases as $F$ increases because of decreased polarization dephasing during the writing and read-out stages. Consequently, near-unity memory efficiency is possible with $F\gg1$ and $\tilde{d}\gtrsim 3$ for the backward-propagating signal mode. It is worth emphasizing that although $F\approx 1$ and $F<1$ are unfavorable ATS regimes (but  favorable EIT regimes), the basic principle of our protocol will still work in a robust manner, but with lower storage efficiency.

Finally, we note that in the near-optimal effective optical depth regime, the maximum memory efficiency for a given $\Gamma$ and signal field with $B_{\rm FWHM}$ is achieved only by control pulse areas $A_{\rm c}^{\rm write} = A_{\rm c}^{\rm read} = 2\pi$. However, this requirement is relaxed to some extent in the non-optimal effective optical depth regime, as can be inspected from Eq.~\ref{eff2} in conjuction with Eqs.~\ref{eq:effForward1} and \ref{eq:effBack1}. When $\tilde{d}$ is significantly smaller than 1.5, a pulse area $A_{c}<2\pi$ will lead to larger efficiency, as it effectively increases $\tilde{d}$ and yields larger $\mu_{\rm abs}\mu_{\rm re}$. However, at the same time, $A_{\rm c}<2\pi$  results in smaller $\mu_{\rm w}\mu_{\rm r}$. With this trade-off, the optimal efficiency can be found with pulse areas as small as $A_{\rm c} = 1.6\pi$. Similarly, when $\tilde{d}$ is significantly larger than 1.5 for the forward-propagating mode, the memory efficiency can be optimized by decreasing $\tilde{d}$ with pulse areas $A_{\rm c}>2\pi$. This leads to a compromise between the increase of $\mu_{\rm abs}\mu_{\rm re}$ and decrease of $\mu_{\rm w}\mu_{\rm r}$ such that the efficiency can be maximized away from $A_{\rm c}=2\pi$. In contrast, for the backward propagating-mode in the large optical depth regime, the maximum efficiency is always obtained by the areas of $A_{\rm c}=2\pi$, as larger optical depths are always favourable.

\subsection*{Cold atom production}
Our experiments are performed in an apparatus designed for (and capable of) producing Bose-Einstein condensates (BECs) of $^{87}$Rb.  For these initial demonstrations of the ATS memory, we use atoms from the first stages of this process, immediately following the initial laser cooling stages of a magneto-optical trap (MOT).  

Our apparatus consists of a diffusive oven of Rb atoms, which are mildly collimated into a first 2-dimensional MOT chamber, where they undergo cooling along two transverse directions.  The resulting atom beam is guided through 15 cm of differential pumping into a final ultrahigh vacuum chamber ($\approx10^{-12}$~Torr) made of a borosilicate glass cell.  Here, the atoms are collected and cooled in a standard six-beam MOT operating with a red-detuning of $\approx3\Gamma$. The laser frequencies used for cooling match the standard D2 transitions $\ket{F = 2}\rightarrow \ket{F^\prime = 3}$ for the laser cooling and $\ket{F = 1}\rightarrow \ket{F^\prime = 2}$ for a ``repump'' beam (Fig.~\ref{fig:rblines}). While the frequency of the repump laser is servo-locked to the $\ket{F = 1}\rightarrow \ket{F^\prime = 2}$ line using saturation absorption spectroscopy, the frequency of the cooling laser is stabilized by beat-note locking to the repump laser. After 15 seconds of laser cooling, we collect $2.0$-$2.5\times10^8$ atoms in $\ket{F=2}$. At this stage, the MOT magnetic field and the repump beams are switched off, but the cooling laser beams are kept on. After a 4~ms wait, which ensures that the mechanical shutter for the repump beam is completely closed, the cooling laser beams are turned off. This process (leading to an off-resonant pumping via the cooling beams) transfers the atoms from $\ket{F=2}$ to $\ket{F=1}$, which serves as the ground level for signal field in the storage experiments. After this transfer, the number of atoms that remains in $\ket{F=2}$ is measured to be only $\sim3-4~\%$ of the initial population. Following an additional 3~ms time-of-flight, which ensures that residual magnetic field from the MOT is further reduced, the cold atoms are then subjected to the memory read and write beams for storing and retrieving a signal field.  Time-of-flight measurements of this cloud yield a size of $3.2$~mm  diameter ($1/e^{1/2}$ diameter from gaussian fit), and temperatures of about 430$~\mu$K. The peak optical depth is determined to be  $ 4$ to $4.5$ from an absorption imaging measurement on $\ket{F = 2}\rightarrow \ket{F^\prime = 3}$ after pumping the atoms back into $\ket{F=2}$. 


Future experiments will test the ATS memory in subsequent stages of cooling these atoms in optical trap and eventually in BEC.  

\subsection*{ATS memory setup and measurements}
In our experiments, the signal ($\ket{F = 1}\rightarrow \ket{F^\prime = 2}$) and control ($\ket{F = 2}\rightarrow \ket{F^\prime = 2}$) fields are derived from the cooling repump beam (an extended cavity diode laser) and a tunable titanium sapphire laser (Fig.~\ref{fig:rblines}), respectively. The frequency of each laser is independently servo-locked to the corresponding D2 transition with a $-80$ MHz detuning using saturation absorption spectroscopy. The continuous-wave (cw) signal field is first gated by an acousto-optic modulator (AOM) that introduces a  frequency shift of $-80$~MHz. Then, it is sent to another AOM in double-pass configuration that generates resonant signal pulses with $\tau_{\rm FWHM}=30-70$~ns by shifting the frequency $+160$~MHz. This AOM is also used for ms-long frequency-swept signal pulses with a $\pm20$~MHz scan range to spectrally characterize ATS. The cw control field is passed through an AOM that produces resonant write and read-out pulses with 30 to 70 ns duration at FWHM and a $+80$~MHz frequency shift. In the memory setup, the AOMs for the signal and control pulses are driven by amplified RF oscillators that are amplitude controlled by a two-channel arbitary-waveform generator (AWG) with a 250 MS/s sampling rate. In the interference measurements, the signal pulses are instead generated using a single-pass AOM with detuning of $+80$~MHz, and both the signal and control AOMs are directly driven by AWG with larger sampling rate (4 GS/s) that allows preparation of signal pulses with the desired phase.  

After setting the frequency, phase, and amplitude of the signal and control fields, each beam is separately coupled to a polarization-maintaining fiber with an output at the cold-atom apparatus. The beams are launched toward the 3D-MOT chamber with fiber-decoupling lenses that collimate the beams with mm-scale waists. The polarization of the each beam (initially linear) is controlled by quarter-wave (QWP) plates, for providing the maximum interaction strength between the fields and atoms. Following the QWPs, the signal and control beams  are combined on a $50/50$ beamsplitter and overlapped inside the atomic cloud with gaussian beam diamaeters ($1/e^2$) of $2.3$~mm and $3.0$~mm and the separation angle of $2^{\circ}$, respectively. The peak power of the signal before the cloud is typically $30~\mu$W for storage experiments and less than $100$~nW for ATS characterization. The peak power of the control field before the cloud is between $0.5$~mW and $15$~mW, depending on the experiment. After interacting with the cold atoms, the signal is coupled to a single-mode fiber for an additional spatial filtering from the strong control field. This yields, in total, $65-70$~dB isolation, which provides no measurable leak from the control to the signal in the storage experiments. Finally, the signal is directed to a fast photo-detector (Resolved Instruments DPD80, $100$~MHz bandwidth) or a slow-detector with large sensitivity ($4$~MHz bandwidth)  for storage or spectral measurements, respectively. In addition to the signal, a small fraction of the control beam that is reflected off a $5/95$ beam splitter is coupled to a fiber and detected to enable  synchronization with the signal, as well as to calibrate the power.

For spectral measurements to characterize the natural line and ATS lines of the ($\ket{F = 1}\rightarrow \ket{F^\prime = 2}$) transition, a 1~ms frequency-swept signal with power of $\mathcal{P}_{\rm in}(\Delta)$ and linearly varying detuning ($\Delta(t)$ from -20~MHz to +20~MHz) is sent to the atomic cloud, and the transmitted signal $\mathcal{P}_{\rm out}(\Delta)$ is detected to determine the absorption with respect to $\Delta$ in terms of optical depth: $d^\prime=\ln(\mathcal{P}_{\rm in}/\mathcal{P}_{\rm out}$). The measured absorption profiles  allow us to directly determine the transition linewidth $\Gamma$ (for $\Omega_{\rm c}=0$) and ATS spacing $\delta_{\rm A}\approx\Omega_{\rm c}$ (for $\Omega_{\rm c}>\Gamma$). This is achieved by fitting each absorption peak trace to a Lorentizan function. Although this measurement should also determine the peak optical depth (at $\Delta=0$), it has yielded $d^{\rm exp}=2.3-2.7$ for the natural transition line, which is about $25\%$ less optical depth compared to an estimation from the imaging measurement, due to a partial destruction of the cloud during interaction. When the sweep range is changed to the detuning range -4~MHz to +36~MHz, this degradation is significantly less, yielding peak optical depth of $d^{\rm exp}\approx 3.1$. To further confirm, the natural transition line is probed with a resonant pulse with $\sim 0.5~\mu$s duration and $30~\mu$W (yielding less power delivery per spectral interval). The measured peak optical depth is $d^{\rm exp}\approx3.3$ with some fluctuations due to laser frequency jitter. 

In a typical storage experiment, a short signal pulse and a write field are simultaneously sent to the atomic cloud. After a pre-set storage time controlled by the AWG, a read-out field is generated for retrieval. The detected time series of the retrieved and  transmitted (non-absorbed part of the input)  signals are recorded on an oscilloscope. To avoid potential degradation to the cloud by several signal pulses, only one signal pulse is stored and retrieved per cooling cycle. This measurement cycle is typically repeated 3-5 times and the data is averaged to improve the signal-to-noise ratio. The area under the curve of power vs.\ time is determined for the output (retrieved) signal and compared to the input signal (measured in the absence of atoms), using gaussian fits to the data (which are generally well fit). The ATS memory efficiency is determined from the ratio of the retrieved area to the input area. For a given memory configuration (with certain bandwidth and storage time), the memory efficiency exhibited variations up to $\pm15~\%$  due to instabilities in atom number (optical depth) and spin-wave decoherence time, and  up to $\pm10~\%$ and $\pm5~\%$ drift of the  power of input signal and control fields, respectively, on the minute-to-minute timescale. 

In our experiments, the interference measurements demand a fixed phase relationship between the control and signal fields, which is not guaranteed by the  measurement procedure described above, where each field is derived from independent lasers and also carried over meters-long fiber-optic patchcords. To overcome this limitation, we send a stream of pulse quartets -- an input signal pulse (each encoded with different relative phase), a reference signal pulse (to be interfered with each recalled signal), and two control pulses (each with the same phase) -- to the atomic cloud to effect  storage and retrieval.  Several of these quartets are sent in less than $5~\mu$s, with varying input signal pulse phases in each quartet, which ensures that there is no significant phase drift between each storage/recall due to the short time scale. We confirmed that each signal pulse experiences the same absorption, by observing that no measurable destruction of the cloud is observed until the last pulse quartet. Furthermore, in order to avoid any cross-talk, additional read-out pulses are applied between each storage/recall quartet, guaranteeing that any spin coherence left from the storage of the previous input and reference signal is removed for the next interference measurement. 

Before the measurements detailed above, a number of calibration, alignment and optimization procedures were performed~\cite{SI}.

\subsection*{Performance of our ATS memory implementation}
In our proof-of-principle demonstrations of an ATS memory, our typical sequence uses an input signal with bandwidth $B_{\rm FWHM}=11$~MHz ($\tau_{\rm FWHM}=40$~ns). This signal is stored for $T=200$~ns and retrieved with $\eta^{\exp}=7-8\%$ efficiency in the forward direction. By adapting Eq.~\ref{eq:effForward2} to our experimental configuration, the memory efficiency is determined by $d_{\rm A}$ (peak optical depth of ATS lines), $F_{\rm eff}$ (effective ATS factor that takes into account additional broadening in the ATS lines), and decay time constant $T_{\rm d}$ for a given storage time $T$ due to spin-wave decoherence,
\begin{align}
\eta^{\rm exp}\approx{(d_{\rm A}/2F_{\rm  eff})}^2 e^{-d_{\rm A}/2F_{\rm eff}} e^{-1/F_{\rm eff}} e^{-T/T_{\rm d}}.
\label{eff5}
\end{align}

We determine $T_{\rm d}$ experimentally by extending the storage time up to 1.0~$\mu$s and measuring the memory efficiency for different storage times in this range~\cite{SI}.
From measurements performed at different times,  we find that $T_{\rm d}$  is typically ($330\pm 15$)~ns, but can be as low as ($220\pm 15$)~ns depending on the bias magnetic field optimization. These measurements also  establish the spin decoherence rates in our system  as $\gamma_{\rm s}^{\rm exp}/2\pi=1/2\pi(2T_{\rm d})\approx 0.24$ to 0.4~MHz. In the ideal case ($\gamma_{s}=0$ and no extra ATS line broadening), as treated in Eq.~\ref{eq:effForward2}, the peak optical depth of the ATS lines is the same as the measured peak optical depth of the transition line ($d_{\rm A}=d^{\rm exp}$), and the width of each ATS line is $\Gamma_{\rm ATS}=\Gamma^{\rm exp}/2$, which defines $F_{\rm eff}=\Omega_{\rm c}/(2\Gamma_{\rm ATS})=\Omega_{\rm c}/\Gamma^{\rm exp}$. In practical settings,  when the spin-decoherence rate ($\gamma_{\rm s}$) is comparable to the polarization decay rate ($\gamma_{\rm e}$) and/or the control Rabi frequency (which determines the ATS splitting) is not uniform across the medium, then the ATS peaks become wider and lower. This effect can be analyzed using an experimentally obtained ATS spectrum, which is shown in Fig.~\ref{setup}c for our typical setting $\Omega_{\rm c}/2\pi \approx 11$ MHz. From this spectrum we measure the linewidth of each ATS peak at FWHM is to be $\Gamma_{\rm ATS}^{\rm exp}/2\pi\approx 1.5\times(\Gamma^{\rm exp}/2)/2\pi\approx6.0$ MHz for $\Gamma^{\rm exp}/2\pi\approx7.7$ MHz (the measured transition linewidth including residual Doppler broadening), which yields $F_{\rm eff}=F/1.5\approx 1$. Similarly we find that the peak optical depth is reduced with approximately the same factor, yielding $d_{\rm A}\approx d^{\rm exp}/1.5\approx2.3$ for our measured peak optical depth $d^{\rm exp}=3.5$. 

We further investigate the origin of the observed line broadening of $\delta \Gamma_{\rm ATS}/2\pi=(\Gamma_{\rm ATS}^{\rm exp}-0.5\times\Gamma^{\rm exp})/2\pi\approx 2$ MHz for $\Omega_{\rm c}/2\pi \approx11$~MHz. In order to find the contribution from the spin decoherence effect,  we simulate the absorption profiles of the ATS system using the generic susceptibility expression for a three-level system for the measured spin-decoherence rates (0.24 to 0.35~MHz) and the control Rabi frequency ranges ($\Omega_{\rm c}/2\pi\leq 17 $ MHz) used in the experiments. We establish this contribution to be in the range of $\approx 0.20-0.30$ MHz, which is, in most cases, neglibly small due to the fact that $\gamma_{s}\ll\gamma_{e}$ is satisfied in our experiments. Therefore, the major contribution to the ATS broadening comes from the spatial non-uniformity of the control Rabi frequency over the interaction cross-section of the atomic gas, which leads to position-dependent ATS splitting that manifests as broadening in their spectra. In our experiment, this non-uniformity originates from the similar beam sizes of the overlapping control and signal fields, each with gaussion intensity profiles in diamaters of 3.0 and 2.3 mm, respectively. To further confirm, we characterize the degree of the ATS line broadening as a function of the power of the control field. We observe that the broadening increases proportionally to the square root of the power ($\delta\Gamma_{\rm ATS}\propto\sqrt{P_{\rm c}}$), and  for relatively high power values ($P_{\rm c}>5$ mW) the spectrum of each ATS line is predominantly characterized by a gaussian spectral profile, indicating the inhomogeneous nature of the broadening due to varying ATS splittings. In the general treatment of a dynamic ATS including this effect, the power dependence of the broadening as well as the alteration in the ATS spectra, which accordingly modifies the definitions of $F$ and $\tilde{d}$, need to be taken into account. For the cases that $\delta\Gamma_{\rm ATS}/2\ll\Omega_{c}$, as in our implementation whose overall efficiency has degraded about $30\%$ due to this broadening effect, treatment of an effective line broadening (reducing $F$) with nearly preserved effective optical depth is a reasonable approximation, as the system dynamics are still mainly governed by the mean $\Omega_{\rm c}$.

Consequently, using the experimentally extracted parameters of $d_{\rm A}=2.3$, $F_{\rm eff}=1$ and $T_{\rm d}=300$~ns in Eq.~\ref{eff5} gives a predicted memory efficiency of $8\%$ which is in agreement with the directly measured memory efficiency of $\eta^{\exp}=7-8\%$.

Beyond our proof-of-principle demonstration, there are clear paths to substantially improve the performance of our ATS memory implementation in terms of efficiency, storage time, and bandwidth~\cite{SI}.

\section*{Acknowledgments}
We appreciate generous technical support from Greg Popowich and Scott Wilson, and the following groups for lending us equipment for our initial measurements: J. Beamish, J. P. Davis,  F. Hegmann, A. Lvovsky, W. Tittel, R. Wolkow. We also thank Dr. Barry Sanders, Dr.Ying-Cheng Chen, Dr. Chris O'Brien for useful discussions.  We gratefully acknowledge funding from the Natural Science and Engineering Research Council of Canada (NSERC RGPIN-2014-06618), Canada Foundation for Innovation (CFI), Canada Research Chairs Program (CRC), Canadian Institute for Advanced Research (CIFAR), Alberta Innovates - Technology Futures (AITF), and the University of Alberta.

\section*{Author contributions}
The ATS memory approach was proposed by E.S. with some feedback from K.H. and L.J.L. The project was supervised by L.J.L. and E.S. The ultracold atom apparatus was designed by  L.J.L., and it was built and commissioned by L.J.L., T.H. and E.S. The design of the experiments, the measurements and the analysis of the results were performed by E.S. and T.H. The numerical modelling of the  ATS memory was performed by K.H. with input from E.S. The simulations and numerical analysis were performed by E.S. and A.R. with guidance of K.H. The manuscript was written by E.S. and L.J.L. with feedback from all co-authors. 

\section*{Additional information}
The authors declare that they have no competing financial interests.

Correspondence and requests for materials should be addressed to Lindsay J. LeBlanc (email: \mbox{lindsay.leblanc@ualberta.ca)} or Erhan Saglamyurek (email: \mbox{saglamyu@ualberta.ca).}

 

\begin{thebibliography}{10}
\expandafter\ifx\csname url\endcsname\relax
  \def\url#1{\texttt{#1}}\fi
\expandafter\ifx\csname urlprefix\endcsname\relax\def\urlprefix{URL }\fi
\providecommand{\bibinfo}[2]{#2}
\providecommand{\eprint}[2][]{\url{#2}}

\bibitem{Autler:1955gb}
\bibinfo{author}{Autler, S.~H.} \& \bibinfo{author}{Townes, C.~H.}
\newblock \bibinfo{title}{{Stark Effect in Rapidly Varying Fields}}.
\newblock \emph{\bibinfo{journal}{Phys. Rev.}} \textbf{\bibinfo{volume}{100}},
  \bibinfo{pages}{703--722} (\bibinfo{year}{1955}).

\bibitem{Picque:1976dh}
\bibinfo{author}{Picque, J.~L.} \& \bibinfo{author}{Pinard, J.}
\newblock \bibinfo{title}{{Direct observation of the Autler-Townes effect in
  the optical range}}.
\newblock \emph{\bibinfo{journal}{Journal of Physics B: Atomic and Molecular
  {\ldots}}}  (\bibinfo{year}{1976}).

\bibitem{He:1992dx}
\bibinfo{author}{He, X.~F.}, \bibinfo{author}{Fisk, P. T.~H.} \&
  \bibinfo{author}{Manson, N.~B.}
\newblock \bibinfo{title}{{Autler{\textendash}Townes effect of the photoexcited
  diamond nitrogen-vacancy center in its triplet ground state}}.
\newblock \emph{\bibinfo{journal}{J. Appl. Phys.}}
  \textbf{\bibinfo{volume}{72}}, \bibinfo{pages}{211--217}
  (\bibinfo{year}{1992}).

\bibitem{Zhu:1990jr}
\bibinfo{author}{Zhu, Y.~F.} \emph{et~al.}
\newblock \bibinfo{title}{{Vacuum Rabi Splitting as a Feature of
  Linear-Dispersion Theory - Analysis and Experimental-Observations}}.
\newblock \emph{\bibinfo{journal}{Phys. Rev. Lett.}}
  \textbf{\bibinfo{volume}{64}}, \bibinfo{pages}{2499--2502}
  (\bibinfo{year}{1990}).

\bibitem{Thompson:1992jr}
\bibinfo{author}{Thompson, R.~J.}, \bibinfo{author}{Rempe, G.} \&
  \bibinfo{author}{Kimble, H.~J.}
\newblock \bibinfo{title}{{Observation of Normal-Mode Splitting for an Atom in
  an Optical Cavity}}.
\newblock \emph{\bibinfo{journal}{Phys. Rev. Lett.}}
  \textbf{\bibinfo{volume}{68}}, \bibinfo{pages}{1132--1135}
  (\bibinfo{year}{1992}).

\bibitem{Bernadot:1992kg}
\bibinfo{author}{Bernadot, F.}, \bibinfo{author}{Nussenzveig, P.},
  \bibinfo{author}{Brune, M.}, \bibinfo{author}{Raimond, J.~M.} \&
  \bibinfo{author}{Haroche, S.}
\newblock \bibinfo{title}{{Vacuum Rabi Splitting Observed on a Microscopic
  Atomic Sample in a Microwave Cavity}}.
\newblock \emph{\bibinfo{journal}{EPL}} \textbf{\bibinfo{volume}{17}},
  \bibinfo{pages}{33--38} (\bibinfo{year}{1992}).

\bibitem{Wade:2014} Wade, C. G. et al. Probing an excited-state atomic transition using hyperfine quantum-beat spectroscopy. \textit{Phys. Rev. A} \textbf{90}, 3 (2014).

\bibitem{Holloway:2014} Holloway, C. L. et al. Sub-wavelength imaging and field mapping via electromagnetically induced transparency and Autler-Townes splitting in Rydberg atoms. \textit{App. Phys. Lett.} \textbf{104}, 24 (2014).

\bibitem{Ghafoor:2014} Ghafoor, F. Autler--Townes multiplet spectroscopy. \textit{Laser Phys.} \textbf{24}, 3 (2014).  

\bibitem{Lukin:2003ct}
\bibinfo{author}{Lukin, M.~D.}
\newblock \bibinfo{title}{{Colloquium: Trapping and manipulating photon states
  in atomic ensembles}}.
\newblock \emph{\bibinfo{journal}{Rev. Mod. Phys.}}
  \textbf{\bibinfo{volume}{75}}, \bibinfo{pages}{457--472}
  (\bibinfo{year}{2003}).

\bibitem{Fleischhauer:2005da}
\bibinfo{author}{Fleischhauer, M.}, \bibinfo{author}{Imamoglu, A.} \&
  \bibinfo{author}{Marangos, J.~P.}
\newblock \bibinfo{title}{{Electromagnetically induced transparency: Optics in
  coherent media}}.
\newblock \emph{\bibinfo{journal}{Rev. Mod. Phys.}}
  \textbf{\bibinfo{volume}{77}}, \bibinfo{pages}{633--673}
  (\bibinfo{year}{2005}).

\bibitem{AbiSalloum:2010eg}
\bibinfo{author}{Abi-Salloum, T.~Y.}
\newblock \bibinfo{title}{{Electromagnetically induced transparency and
  Autler-Townes splitting: Two similar but distinct phenomena in two categories
  of three-level atomic systems}}.
\newblock \emph{\bibinfo{journal}{Phys. Rev. A}} \textbf{\bibinfo{volume}{81}},
  \bibinfo{pages}{053836--6} (\bibinfo{year}{2010}).

\bibitem{Anisimov:2011fnb}
\bibinfo{author}{Anisimov, P.~M.}, \bibinfo{author}{Dowling, J.~P.} \&
  \bibinfo{author}{Sanders, B.~C.}
\newblock \bibinfo{title}{{Objectively Discerning Autler-Townes Splitting from
  Electromagnetically Induced Transparency}}.
\newblock \emph{\bibinfo{journal}{Phys. Rev. Lett.}}
  \textbf{\bibinfo{volume}{107}}, \bibinfo{pages}{163604--4}
  (\bibinfo{year}{2011}).

\bibitem{Giner:2013} Giner, L. et al. Experimental investigation of the transition between Autler-Townes splitting and electromagnetically-induced-transparency models. \textit{Phys. Rev. A} \textbf{87}, 013823 (2013).


\bibitem{Tan:2014hk}
\bibinfo{author}{Tan, C.} \& \bibinfo{author}{Huang, G.}
\newblock \bibinfo{title}{{Crossover from electromagnetically induced
  transparency to Autler{\textendash}Townes splitting in open ladder systems
  with Doppler broadening}}.
\newblock \emph{\bibinfo{journal}{J. Opt. Soc. Am. B}}
  \textbf{\bibinfo{volume}{31}}, \bibinfo{pages}{704--12}
  (\bibinfo{year}{2014}).

\bibitem{Peng:2014gb}
\bibinfo{author}{Peng, B.}, \bibinfo{author}{Chen, W.}, \bibinfo{author}{Nori,
  F.}, \bibinfo{author}{zdemir, S. a. K.~O.} \& \bibinfo{author}{Yang, L.}
\newblock \bibinfo{title}{{What is and what is not electromagnetically induced
  transparency in whispering-gallery microcavities}}.
\newblock \emph{\bibinfo{journal}{Nature Communications}}
  \textbf{\bibinfo{volume}{5}}, \bibinfo{pages}{1--9} (\bibinfo{year}{2014}).

\bibitem{Lu:2015dk}
\bibinfo{author}{Lu, X.} \emph{et~al.}
\newblock \bibinfo{title}{{Transition from Autler{\textendash}Townes splitting
  to electromagnetically induced transparency based on the dynamics of decaying
  dressed states}}.
\newblock \emph{\bibinfo{journal}{J. Phys. B: At. Mol. Opt. Phys.}}
  \textbf{\bibinfo{volume}{48}}, \bibinfo{pages}{055003--9}
  (\bibinfo{year}{2015}).

\bibitem{He:2015ho}
\bibinfo{author}{He, L.-Y.}, \bibinfo{author}{Wang, T.-J.},
  \bibinfo{author}{Gao, Y.-P.}, \bibinfo{author}{Cao, C.} \&
  \bibinfo{author}{Wang, C.}
\newblock \bibinfo{title}{{Discerning electromagnetically induced transparency
  from Autler-Townes splitting in plasmonic waveguide and coupled resonators
  system}}.
\newblock \emph{\bibinfo{journal}{Opt. Express}} \textbf{\bibinfo{volume}{23}},
  \bibinfo{pages}{23817--10} (\bibinfo{year}{2015}).

\bibitem{Liao:2014jr}
\bibinfo{author}{Liao, W.-T.}, \bibinfo{author}{Keitel, C.~H.} \&
  \bibinfo{author}{P{\'a}lffy, A.}
\newblock \bibinfo{title}{{All-Electromagnetic Control of Broadband Quantum
  Excitations Using Gradient Photon Echoes}}.
\newblock \emph{\bibinfo{journal}{Phys. Rev. Lett.}}
  \textbf{\bibinfo{volume}{113}}, \bibinfo{pages}{123602--5}
  (\bibinfo{year}{2014}).

\bibitem{Liu:2001} Liu, C., Dutton, Z., Behroozi, C. H. \& Hau, L. V. Observation of coherent optical information storage
in an atomic medium using halted light pulses. \textit{Nature} \textbf{409}, 490-493 (2001).

\bibitem{Hosseini:2009} Hosseini, M. et al. Coherent optical pulse sequencer for quantum applications.
\textit{Nature} \textbf{461}, 241-245 (2009)

\bibitem{Afzelius:2010a} Afzelius, M. et al. Demonstration of atomic frequency comb memory for light with spin-wave storage. \textit{Phys. Rev. Lett.} \textbf{104}, 040503 (2010).

\bibitem{Reim:2010} Reim, K. F. et al. Towards high-speed optical quantum memories. \textit{Nat. Photon.} \textbf{4}, 218-221 (2010)

\bibitem{Hedges:2010} Hedges, M. P., Longdell J. J., Li Y.,	Sellars, M. J. Efficient quantum memory for light. \textit{Nature} \textbf{465}, 1052–1056 (2010).

\bibitem{Clausen:2011} Clausen, C. et al. Quantum storage of photonic entanglement in a crystal. \textit{Nature} \textbf{469}, 508-512 (2011).

\bibitem{Saglamyurek:2011} Saglamyurek, E. et al. Broadband waveguide quantum memory for entangled photons. \textit{Nature} \textbf{469}, 513-518 (2011).



\bibitem{Mohapatra:2007im}
\bibinfo{author}{Mohapatra, A.~K.}, \bibinfo{author}{Jackson, T.~R.} \&
  \bibinfo{author}{Adams, C.~S.}
\newblock \bibinfo{title}{{Coherent Optical Detection of Highly Excited Rydberg
  States Using Electromagnetically Induced Transparency}}.
\newblock \emph{\bibinfo{journal}{Phys. Rev. Lett.}}
  \textbf{\bibinfo{volume}{98}}, \bibinfo{pages}{2047--4}
  (\bibinfo{year}{2007}).

\bibitem{Pritchard:2010im}
\bibinfo{author}{Pritchard, J.~D.} \emph{et~al.}
\newblock \bibinfo{title}{{Cooperative Atom-Light Interaction in a Blockaded
  Rydberg Ensemble}}.
\newblock \emph{\bibinfo{journal}{Phys. Rev. Lett.}}
  \textbf{\bibinfo{volume}{105}}, \bibinfo{pages}{193603--4}
  (\bibinfo{year}{2010}).

\bibitem{Agarwal:2010dp}
\bibinfo{author}{Agarwal, G.~S.} \& \bibinfo{author}{Huang, S.}
\newblock \bibinfo{title}{{Electromagnetically induced transparency in
  mechanical effects of light}}.
\newblock \emph{\bibinfo{journal}{Phys. Rev. A}} \textbf{\bibinfo{volume}{81}},
  \bibinfo{pages}{041803--4} (\bibinfo{year}{2010}).

\bibitem{Huang:2010gv}
\bibinfo{author}{Huang, J.~Y.} \emph{et~al.}
\newblock \bibinfo{title}{{In Situ Observation of the Electrochemical
  Lithiation of a Single SnO2 Nanowire Electrode}}.
\newblock \emph{\bibinfo{journal}{Science}} \textbf{\bibinfo{volume}{330}},
  \bibinfo{pages}{1515--1520} (\bibinfo{year}{2010}).

\bibitem{Teufel:2011ih}
\bibinfo{author}{Teufel, J.~D.} \emph{et~al.}
\newblock \bibinfo{title}{{Circuit cavity electromechanics in the
  strong-coupling regime}}.
\newblock \emph{\bibinfo{journal}{Nature}} \textbf{\bibinfo{volume}{471}},
  \bibinfo{pages}{204--208} (\bibinfo{year}{2011}).

\bibitem{SafaviNaeini:2011dm}
\bibinfo{author}{Safavi-Naeini, A.~H.} \emph{et~al.}
\newblock \bibinfo{title}{{Electromagnetically induced transparency and slow light with optomechanics}}.
\newblock \emph{\bibinfo{journal}{Nature}} \textbf{\bibinfo{volume}{472}},
  \bibinfo{pages}{69--73} (\bibinfo{year}{2011}).

\bibitem{Sillanpaa:2009db}
\bibinfo{author}{Sillanp{\"a}{\"a}, M.~A.} \emph{et~al.}
\newblock \bibinfo{title}{{Autler-Townes Effect in a Superconducting
  Three-Level System}}.
\newblock \emph{\bibinfo{journal}{Phys. Rev. Lett.}}
  \textbf{\bibinfo{volume}{103}}, \bibinfo{pages}{181--4}
  (\bibinfo{year}{2009}).

\bibitem{Abdumalikov:2010gw}
\bibinfo{author}{Abdumalikov, A.~A.} \emph{et~al.}
\newblock \bibinfo{title}{{Electromagnetically Induced Transparency on a Single
  Artificial Atom}}.
\newblock \emph{\bibinfo{journal}{Phys. Rev. Lett.}}
  \textbf{\bibinfo{volume}{104}}, \bibinfo{pages}{193601--4}
  (\bibinfo{year}{2010}).

\bibitem{Novikov:2013jt}
\bibinfo{author}{Novikov, S.} \emph{et~al.}
\newblock \bibinfo{title}{{Autler-Townes splitting in a three-dimensional
  transmon superconducting qubit}}.
\newblock \emph{\bibinfo{journal}{Phys. Rev. B}} \textbf{\bibinfo{volume}{88}},
  \bibinfo{pages}{645--5} (\bibinfo{year}{2013}).

\bibitem{Sun:2014kh}
\bibinfo{author}{Sun, H.-C.} \emph{et~al.}
\newblock \bibinfo{title}{{Electromagnetically induced transparency and
  Autler-Townes splitting in superconducting flux quantum circuits}}.
\newblock \emph{\bibinfo{journal}{Phys. Rev. A}} \textbf{\bibinfo{volume}{89}},
  \bibinfo{pages}{1071--17} (\bibinfo{year}{2014}).


\bibitem{Gorshkov:2007gd}
\bibinfo{author}{Gorshkov, A.~V.}, \bibinfo{author}{Andr{\'e}, A.},
  \bibinfo{author}{Lukin, M.~D.} \& \bibinfo{author}{S{\o}rensen, A.~S.}
\newblock \bibinfo{title}{{Photon storage in $\Lambda$-type optically dense
  atomic media. II. Free-space model}}.
\newblock \emph{\bibinfo{journal}{Phys. Rev. A}} \textbf{\bibinfo{volume}{76}},
  \bibinfo{pages}{033804--25} (\bibinfo{year}{2007}).



\bibitem{Afzelius:2009} Afzelius, M., Simon, C., de Riedmatten H., Gisin, N. Multimode quantum memory based on atomic frequency combs. \textit{Phys. Rev. A} \textbf{79}, 052329 (2009)

\bibitem{Tittel:2010} Tittel, W. et al. Photon-echo quantum memory in solid state system. \textit{Laser \& Photon. Rev.}  \textbf{4}, No. 2, 244–267 (2010)

\bibitem{Riedll:2012} Riedl, S. et al. Bose-Einstein condensate as a quantum memory for a photonic polarization qubit. \textit{Phys. Rev. A} \textbf{85}, 022318 (2012).

\bibitem{Choo:2016} Cho, Y-W. et al.Highly efficient optical quantum memory with long coherence time in cold atoms. \textit{Optica} \textbf{3}, 100-107 (2016)

\bibitem{Lukin:2001} Lukin, M. D. \& Imamoglu, A. Controlling photons using electromagnetically induced transparency. \textit{Nature} \textbf{413}, 273-276 (2001)

\bibitem{Lvovsky:2009} Lvovsky, A. I., Tittel, W. \& Sanders, B. C. Optical quantum memory. \textit{Nature Photon.} \textbf{ 3}, 706-714 (2009).


\bibitem{Saglamyurek:2014} Saglamyurek, E. et al. An integrated processor for photonic quantum states using a broadband light-matter interface. \textit{New J. Phys.} \textbf{16}, 065019  (2014).


\bibitem{SI} For details, see Supplementary Information. 


\bibitem{Reim:2013} Reim, K. F. et al. Multipulse Addressing of a Raman Quantum Memory: Configurable Beam Splitting and Efficient Readout. \textit{Phys. Rev. Lett.} \textbf{108}, 263602 (2013).

\bibitem{Campbell:2014} Campbell, G. T. et al. Configurable Unitary Transformations and Linear Logic Gates Using Quantum Memories. \textit{Phys. Rev. Lett.} \textbf{113}, 063601 (2014).

\bibitem{Heshami:2016} Heshami, K. et al. Quantum memories: emerging applications and recent advances. \textit{J. Mod. Opt.} \textbf{63}, 20 (2016).

\bibitem{Kuzmich} Lan,  S. Y. et al., A Multiplexed Quantum Memory. \textit{Opt. Express} {\bf 17}, 13639 (2009).

\bibitem{Pu:2017} Pu, Y-F et al. Experimental realization of a multiplexed quantum memory with 225 individually accessible memory cells. \textit{Nature Commun.} \textbf{8},15359 (2017).


\end{thebibliography}

\begin{thebibliography}{1} 
 
 \bibitem{Cho:2016} Cho, Y-W. et al.Highly efficient optical quantum memory with long coherence time in cold atoms. \textit{Optica} \textbf{3}, 100-107 (2016)

\end{thebibliography}

\pagebreak
\begin{widetext}

\begin{center}
\textbf{\large SUPPLEMENTARY INFORMATION}
\end{center}
\end{widetext}

\setcounter{equation}{0}
\setcounter{figure}{0}
\setcounter{table}{0}
\setcounter{page}{1}
\makeatletter
\renewcommand{\theequation}{S\arabic{equation}}
\renewcommand{\thefigure}{S\arabic{figure}}
\renewcommand{\thetable}{S\arabic{table}}
\renewcommand{\thesection}{S\Roman{section}}
\renewcommand{\thesubsection}{s\roman{subsection}}

\section{Optimization, calibration and alignment procedures for experiments}

We follow various preparation procedures for our experiments as detailed below. First, the positions of the signal and control beams are aligned with respect to the atomic cloud using strong resonant fields that remove atoms from the interacting section of the cloud. This section is monitored using absorption imaging on a CCD camera, and the beams are aligned until a reasonable overlap is obtained. Second, the control field peak power is regularly calibrated by comparing a set of actual power measurements before the atomic cloud with the ones measured from the monitor detector. Third, the polarization of the control and signal beams is optimized by maximizing  both peak absorption and peak spacing of the ATS $\delta_{\rm A}$ (measured with the transmission of a frequency-swept signal) for a given control field power. Fourth, for the pulsed-ATS memory demonstrations, the synchronization between the signal and the control pulses was carried out by monitoring signals from the respective detectors, which are positioned with equal optical distance from the atomic cloud. When required, the relative delay is tuned by introducing a trigger delay to the AWG channels. Fifth, in order to optimize spin coherence time, the residual magnetic field at the cloud location is compensated to near-zero using three pairs of bias coils that can independently produce uniform DC magnetic fields in each of the three dimensions. This procedure was carried out by maximizing the efficiency of the  memory with iterative adjustments of the bias magnetic fields. Finally, atom number (optical depth) is optimized day-to day by optimizing laser-cooling parameters.

\section{Limitations and potential improvements in our ATS memory implementation}

The main limitation in our experiments is a fast spin-decoherence rate, which degrades the efficiency and the storage time. This limitation is mainly due to residual magnetic fields, leading to dephasing between the atoms that are populating different Zeeman sublevels. The slow switch-off time of the MOT coils is one  source of this residual field, which is not totally compensated by the DC-bias fields. One way to overcome this issue in the current setup is to implement polarization-gradient cooling (optical molasses), which would extend the time after switching off the MOT fields before storage and retrieval, and would additionally increase the optical density. In conjunction with microwave or RF spectroscopy, which systematically allows the measurement and, therefore, cancellation of residual fields, it should be possible to optimize the spin-coherence time from the current value of a few-hundred nanoseconds to a few-hundred microseconds, as demonstrated in several experiments with cold atoms~\cite{Cho:2016}. The other factor limiting the efficiency is the small optical depths, which can be  significantly increased with polarization gradient and evaporative cooling. Moreover, as our apparatus is capable of producing ultracold atoms in an optical dipole trap,  more than an order of magnitude improvement in optical depth is within our reach. In combination with sufficiently large coherence times, this improvement should lead to the memory efficiency larger than $50\%$ with the backward-propagation scheme. Finally, in our current setup, the memory bandwidth is technically limited by our electronics and AOMs, which give the shortest  pulses of $\tau_{\rm FWHM}=30$ ns ($B_{\rm FWHM}=14$ MHz), and by the peak Rabi frequency $\Omega_{\rm c}/2\pi\approx20$ MHz. As the Rabi frequency depends on the power and beam diameter, reducing the beam diameter from the current $3$~mm to $0.3$~mm (with the corresponding  change of the signal beam size), we can obtain Rabi frequencies of $200$~MHz, which, in principle, should allow $200$~MHz memory bandwidth. On the other hand, as the spacing between the adjacent D2 excited levels is only on the order of 100 MHz, such a broadband memory scheme would not be possible with D2. An alternative solution to this limitation is  to employ the D1 line in $^{87}$Rb which has about 800~MHz spacing between its excited levels.  Together with sufficiently large optical depths, such an implementation should allow a few-hundred MHz bandwidth in our current setup.

\begin{figure}[tb]
\begin{center}
\includegraphics{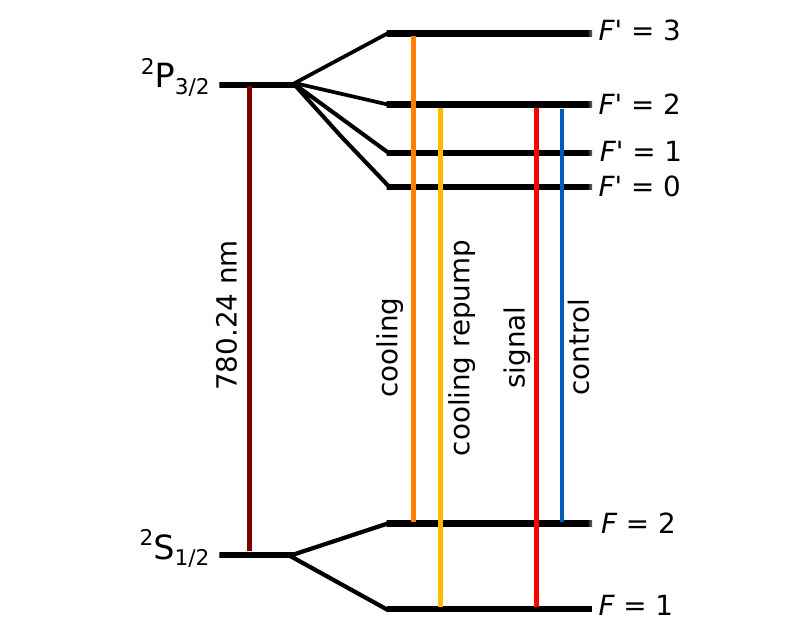}
\caption{\textbf{Rubidium 87 level structure.} Relevant laser transitions labelled, left-to-right: Bare D2 transition is 780.24 nm; cooling transition used for MOT; repump cooling transition used for MOT; signal transition $\ket{F = 1}\rightarrow \ket{F^\prime = 2}$; control transition $\ket{F = 2}\rightarrow \ket{F^\prime = 2}$}
\label{fig:rblines}
\end{center}
\end{figure}

\begin{figure}[tb]
\begin{center}
\includegraphics{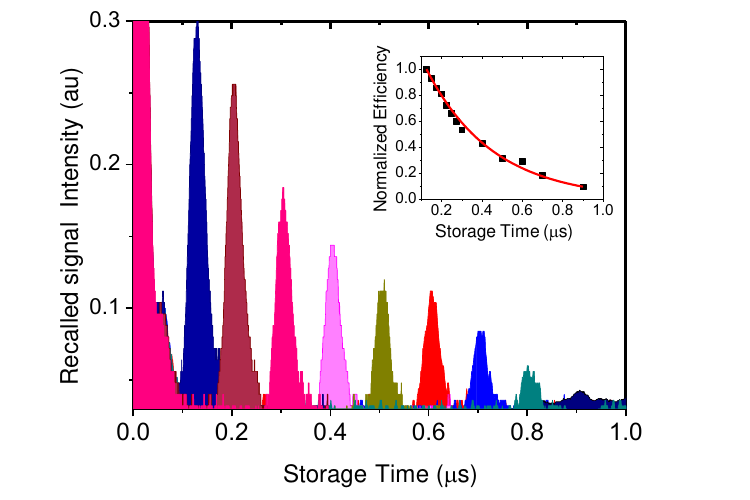}
\caption{\textbf{Retrieved signal intensity with respect to storage time} An input signal with $\tau=40$ ns is stored and retrieved up to a microsecond. The decay arises from spin-wave decoherence, whose main source is  non-zero ambient magnetic fields that lift degeneracy of the Zeeman levels and lead to dephasing between the spin states. The inset shows a fitting of the normalized memory efficiency to a decaying exponential function, which yields decay time of $(330\pm13)$ ns }
\label{fig:decay}
\end{center}
\end{figure}

\section{Numerical analyses and simulations of experiments}
The numerical analyses of the Maxwell-Bloch equations in the ATS regime are performed with a standard Euler method using  MATLAB software. Before calculations, the consistency of the atom-light parameters ($d$, $\gamma_{\rm e}$ and $\Omega_{\rm c}$) with respect to their definitions is inspected. First, by considering $\Omega_{\rm c}=0$, the response of the two-level system ($\ket{g}$ and $\ket{e}$) is probed by a test signal pulse for an example setting of $\Gamma^{\rm set}=2\gamma_{\rm e}^{\rm set}$ and $d^{\rm set}$. The bandwidth of the test signal is chosen to be much smaller than $\Gamma^{\rm set}$. Under these conditions, the output signal arising from a zero-detuning input signal is numerically calculated. From the attenuation of the test signal,the  optical depth is found to be $d=d^{\rm set}$, exactly as defined. In addition, this examination is performed for test signals with detunings from $-4\Gamma$  to $+4\Gamma$. For each detuning, the output is calculated, and absorption is plotted with respect to detuning. The resulting Lorentzian curve has a linewidth of  $\Gamma/2\pi=\Gamma^{\rm set}/2\pi$ at FWHM, exactly as defined. Second, the dynamics of the three-level system is  inspected for a test control field in the regime $\Omega_{\rm c}>\Gamma$. In this case, the input test signal's bandwidth is chosen to be larger than $\Omega_{\rm c}^{\rm set}$, and the time dependence of spin, photon and polarization are calculated. As expected from the theory, it is verified that each component oscillates exactly at the period of $\Omega_{\rm c}/2\pi=\Omega_{\rm c}^{\rm set}/2\pi$.

The simulation of the storage experiments is performed  using the measured atom-light parameters $d^{\rm exp}$, $\Gamma^{\rm exp}$ and $\Omega_{\rm c}$ (extracted from independent experiments), and the recorded input and control field traces, which contain all timing information of the experiment and control power. The Rabi frequency for the control field is calibrated according to $\Omega_{\rm c}(t)/2\pi=\alpha\sqrt{\mathcal{P}(t)}$ with $\alpha=5.75$ MHz/$\sqrt{{\rm mW}}$, which is established from the ATS vs.\ control power characterizations. In addition, the non-uniformity of the Rabi frequency, resulting in broadened ATS lines ($\delta\Gamma_{\rm ATS}$) needs to be considered, otherwise the numerically calculated memory efficiency is approximately $25-30\%$ more than the directly measured one, while the dynamics are unchanged. This effect is  taken into account with a simple model based on our experimental conditions, including $\gamma_{\rm s}\ll\gamma_{\rm e}$, $\delta\Gamma_{\rm ATS}/2\ll\Omega_{\rm c}$ and $B_{\rm FWHM}\approx\Omega_{\rm c}/2\pi$. As detailed in Methods, under these conditions the overall impact of the ATS line broadening is to reduce the memory efficiency by inducing a larger polarization decay rate (i.e. decreasing $F$), while the effective optical depth ($\tilde{d}$) is approximately preserved. Also by considering that the measured transition width ($\Gamma^{\rm exp}$) includes a residual Doppler broadening, we define the atom-light coupling rate, the effective polarization decay rate $\gamma_{e}^{\rm eff}$, and the effective peak optical depth $d^{\rm eff}$ as $g\sqrt{N}=\sqrt{{r d^{\rm eff} \gamma_{\rm e}}/{2L}}$,  $\gamma_{\rm e}^{\rm eff}=(\Gamma^{\rm exp}+2\delta \Gamma_{\rm ATS})/2$, and $d^{\rm eff}=d^{\rm exp}(0.5\times\Gamma^{\rm exp})/\gamma^{\rm eff}$ where $r=\gamma_{\rm e}^{\rm eff}/\gamma_{\rm e}$, and  $\delta\Gamma_{\rm ATS}=\Gamma_{\rm ATS}^{\rm exp}-(\Gamma^{\rm exp}/2)$ is the effective ATS line broadening (excluding the residual contribution from the spin-decoherence), which is extracted from the measured ATS spectra. 

As a consistency check,  we compare the results of basic experiments configured for $\Omega_{\rm c} = 0$ (where there is no ATS memory) to numerical simulations with these parameters. In the first experiments, a long pulse with $B_{\rm FWHM}\ll\Gamma^{\rm exp}/2\pi$ is sent to the cold atoms and the transmitted signal is recorded. These experiments are well simulated in terms of both the profile and amplitude of the transmitted signal using $d^{\rm \exp}\approx3.3$ to $3.5$. In the second experiments, a short pulse with $B_{\rm FWHM}>\Gamma^{\rm exp}/2\pi$ is sent to the cold atoms. In addition to an attenuated transmission, a coherently emitted delayed pulse due to a stimulation effect is recorded. The shape, amplitude and delay between the re-emission and transmission are governed by $d$ and $\gamma_{\rm e}$, allowing their inspection together. These experiments are also well-simulated with setting of $d^{\rm exp}\approx3.3$ and $\Gamma^{\rm exp}/2\pi=2\gamma_{\rm e}^{\rm eff}/2\pi=7.7$~MHz confirming the consistency our set parameters. Finally, for the storage experiments, the  spin-wave decay rate parameter ($\gamma_{\rm s}^{\rm exp}$) is extracted from the memory efficiency vs.\ storage time measurements as described in Methods. The consistency of this  parameter is directly verified by simulating each measurement result from the memory efficiency vs.\ storage time experiment. 

Consequently, taking into account the overall instabilities in atoms numbers (optical depth) and spin-coherence times in the day-to-day operation of our system, we set $d^{\rm exp}=3-3.5$, $\Gamma^{\rm exp}/2\pi=7.7$~MHz, $\gamma_{\rm s}^{\rm exp}/2\pi=0.25-0.40$~MHz, $\delta\Gamma_{\rm ATS}^{\rm exp}/2\pi=1.9$~MHz and $\Omega_{\rm c}/2\pi=5.75~{\rm MHz}/\sqrt{{\rm mW}}\times\sqrt{\mathcal{P}}$  in the simulations of all of the experiments. Finally, we point out that the ATS line broadening due to the non-uniform Rabi frequencies, as observed in our experiments, can be easily eliminated by appropriate selection of the beam sizes, for instance, choosing the control beam size much greater than the probe size.

\end{document}